\documentclass[pra,preprint,superscriptaddress,floatfix,showkeys,showpacs]{revtex4-1}

\usepackage{amsmath,amsfonts,amssymb}
\usepackage{graphicx,color}
\allowdisplaybreaks

\def\beq{\begin{equation}}
\def\eeq{\end{equation}}
\def\beqn{\begin{eqnarray}}
\def\eeqn{\end{eqnarray}}
\def\beqs{\begin{subequations}}
\def\eeqs{\end{subequations}}

\begin{document}

\title{Fragmentation of identical and distinguishable bosons'
pairs and natural geminals of a trapped bosonic mixture}
\author{Ofir E. Alon}
\email{ofir@research.haifa.ac.il}
\affiliation{Department of Mathematics, University of Haifa, Haifa 3498838, Israel}
\affiliation{Haifa Research Center for Theoretical Physics and Astrophysics, University of Haifa,
Haifa 3498838, Israel}

\begin{abstract}
In a mixture of two kinds of identical bosons there are two types of pairs,
identical bosons' pairs, of either species, and pairs of distinguishable bosons. 
In the present work fragmentation of pairs in a trapped mixture of Bose-Einstein condensate is investigated
using a solvable model,
the symmetric harmonic-interaction model for mixtures.
The natural geminals for pairs made of identical or distinguishable bosons
are explicitly contracted by diagonalizing the intra-species and inter-species reduced two-particle density matrices, respectively.
Properties of pairs' fragmentation in the mixture are discussed,
the role of the mixture's center-of-mass and relative center-of-mass coordinates is elucidated,
and a generalization to higher-order reduced density matrices is made.
As a complementary result,
the exact Schmidt decomposition of the wavefunction of the bosonic mixture is constructed.
The entanglement between the two species is governed by the coupling of their individual center-of-mass coordinates,
and it does not vanish at the limit of an infinite number of particles where
any finite-order intra-species and inter-species reduced density matrix per particle is 100\% condensed.
Implications are briefly discussed.
\end{abstract}

\maketitle

\section{Introduction}\label{INTRO}

Condensation and fragmentation are basic and widely-studied concepts of Bose-Einstein condensate
emanating from the properties of the reduced one-particle density matrix \cite{RDM1,RDM2,RDM3,RDM4,RDM5}.
The bosons are said to be condensed if there is a single macroscopic eigenvalue of the
reduced one-particle density matrix \cite{PEN_ONS} and fragmented if there
are two or more such macroscopic eigenvalues \cite{FRAG_REV}. 
These eigenvalues are commonly called natural occupation numbers and
the respective eigenfunctions of the reduced one-particle density matrix are referred to as natural orbitals.
Fragmentation of Bose-Einstein condensate has been investigated, e.g.,
in \cite{FRAG1,FRAG2,FRAG3,FRAG4,FRAG5,FRAG6,FRAG7,FRAG8,FRAG9,FRAG10,FRAG11,FRAG12,
FRAG13,FRAG14,FRAG15,FRAG16,FRAG17,FRAG18,FRAG19,FRAG20}.

The condensation and especially fragmentation of
the reduced two-particle density matrix of interacting identical bosons is less studied, see, e.g., \cite{RDMs_FRAG}.
Here, the analysis of the reduced two-particle density matrix
would determine whether pairs of bosons are condensed or fragmented.
The respective eigenfunctions of the reduced two-particle density matrix
are often called natural geminals.
We note that natural geminals in electronic systems have long been explored, see, e.g.,
\cite{GEM1,GEM2,GEM3,GEM4,GEM5,GEM6,GEM7,GEM8,GEM9,GEM10,GEM11,GEM12}.

Consider now a mixture of two kinds of identical bosons,
which are labeled species $1$ and species $2$.
Mixtures of Bose-Einstein condensate is a highly investigated topic, see, e.g., \cite{MIX1,MIX2,MIX3,MIX4,MIX5,MIX6,MIX7,MIX8,MIX9,MIX10,MIX11,MIX12,MIX13,
MIX14,MIX15,MIX16,MIX17,MIX18,MIX19,MIX20,MIX21,MIX22,MIX23,MIX24,MIX25,MIX26,
MIX27,MIX28,MIX29,MIX30}.  
One may ask,
just like for single-species bosons,
about condensation or fragmentation of each of the species
and how, for instance,
one species is affected by the presence of the other species and vice versa.
To answer this question the intra-species reduced one-particle density matrices
of species $1$ and $2$ are required, i.e.,
analyzing the intra-species occupation numbers and natural orbitals.
Following the above line,
one could also investigate fragmentation of higher-order intra-species reduced density matrices in the mixture.
For instance, to investigate whether pairs of identical bosons,
of either species $1$ or species $2$, are fragmented,
diagonalizing of the intra-species reduced two-particle density matrices is needed.
Summarizing, fragmentation of identical bosons and its manifestation
in higher-order reduced density matrices stem from the properties of intra-species quantities.

But, a mixture of Bose-Einstein condensates
offers a degree-of-freedom or many-particle construction
which do not exist for single-species bosons, namely, inter-species reduced density matrices.
Now, if fragmentation of identical bosons and pairs is defined as the macroscopic occupation
of respective eigenvalues following the diagonalization of intra-species reduced density matrices,
we may analogously define fragmentation of distinguishable bosons' pairs as
macroscopic occupation of the eigenvalues of the inter-species reduced two-particle density matrix.
Obviously, the later is the lowest-order inter-species quantity,
since at least one particle of each species is needed to build an
inter-species entity.

The above discussion defines the goals of the present work which are:
(i) To investigate fragmentation of pairs of identical bosons and establish fragmentation of pairs of 
distinguishable bosons in a mixture of Bose-Einstein condensates;
(ii) To construct the respective natural geminals of the mixture, for identical pairs and for distinguishable pairs;
(iii) To show that fragmentation of distinguishable bosons' pairs in the mixture persists with
higher-order inter-species reduced density matrices;
(iv) To construct the Schmidt decomposition of the mixture's wavefunction and discuss
some of its properties at the limit of an infinite-number of particles where the mixture is $100\%$ condensed;
and
(v) Achieving the first four goals analytically, using an exactly solvable model.

To this end we recruit the harmonic-interaction model for mixtures
\cite{HIM_MIX1,HIM_MIX_ENTANGLE,HIM_MIX2,HIM_MIX_RDM,HIM_MIX_VAR,HIM_MIX_FLOQUET},
or, more precisely here, a symmetric version of which \cite{HIM_MIX_CP}.
The harmonic-interaction model for single-species bosons (and fermions)
has been used extensively in the literature including for investigating properties of Bose-Einstein condensates
\cite{HIM_RDM,HIM_DIAG1,HIM_DIAG2,HIM_JCP,HIM_SCH,HIM1,HIM2,HIM3,HIM4,HIM5,HIM6,HIM7,HIM8,HIM9,HIM10}.
In our work
we build on results obtained and techniques used 
for the reduced density matrices of single-species bosons within the harmonic-interaction model
\cite{HIM_RDM,HIM_DIAG1,HIM_DIAG2,HIM_JCP,HIM_SCH},
and, among others, generalize and extend them for 
the intra-species and particularly the inter-species reduced density matrices of mixtures \cite{HIM_MIX_RDM}.

The structure of the paper is as follows.
In Sec.~\ref{PAIR} we construct and investigate fragmentation of
intra-species and inter-species pair functions in the mixture.
In Sec.~\ref{MORE} we extend the results and explore fragmentation of
pairs of distinguishable pairs.
Furthermore, a complementary result for the Schmidt decomposition
of the mixture's wavefucntion at the limit of an infinite number of particles is offered.
In Sec.~\ref{SUM_OUT} a summary of the results and an outlook of some prospected research topics are provided. 
Finally, appendix \ref{APP} collects for comparison with the mixture 
the details of fragmentation of bosons and pairs in the single-species system.

\section{Intra-species and inter-species natural pair functions}\label{PAIR}

\subsection{The symmetric two-species harmonic-interaction model}\label{PAIR_1}

We consider a mixture of two Bose-Einstein condensates
described by the Hamiltonian of the symmetric two-species harmonic-interaction model
\cite{HIM_MIX_CP,HIM_MIX_RDM}:
\beqn\label{HIM_MIX}
& & \hat H(x_1,\ldots,x_N,y_1,\ldots,x_N) =
\sum_{j=1}^{N} \left( -\frac{1}{2m}\frac{\partial^2}{\partial x_j^2} + \frac{1}{2}m\omega^2 x_j^2 \right)
+ \lambda \sum_{1\le j <k}^{N} \left(x_j-x_k\right)^2 + \nonumber \\ 
& & +
\sum_{j=1}^{N} \left( -\frac{1}{2m}\frac{\partial^2}{\partial y_j^2} + \frac{1}{2}m\omega^2 y_j^2 \right)
+ \lambda \sum_{1\le j <k}^{N} \left(y_j-y_k\right)^2 + \lambda_{12} \sum_{j=1}^{N} \sum_{k=1}^{N} \left(x_j-y_k\right)^2. \
\eeqn
There are $N$ bosons of type $1$, $N$ bosons of type $2$,
and the mass of each boson is $m$.
$\lambda$ is the intra-species interaction strength, either between two bosons of type $1$ or two bosons of type $2$,
and $\lambda_{12}$ is the inter-species interaction strength between type $1$ and type $2$ bosons.
Dimensionality plays no role in the present work hence we work in one spatial dimensions.
$\hbar=1$ is used throughout.
Employing Jacoby coordinates for the mixture and translating back to the laboratory frame,
the $2N$-boson wavefunction and corresponding many-particle density matrix are given by
\beqs\label{HIM_MIX_WF_DEN_MAT}
\beqn\label{HIM_MIX_WF_DEN_MAT1}
& &
\Psi(x_1,\ldots,x_N,y_1,\ldots,y_N) =
\left(\frac{m\Omega}{\pi}\right)^{\frac{N-1}{2}}
\left(\frac{M_{12}\Omega_{12}}{\pi}\right)^{\frac{1}{4}}
\left(\frac{M\omega}{\pi}\right)^{\frac{1}{4}} \times \nonumber \\
& &
\times e^{-\frac{\alpha}{2}\sum_{j=1}^N \left(x_j^2 + y_j^2\right) - \beta \sum_{1\le j < k}^N \left(x_jx_k+y_jy_k\right)
+ \gamma \sum_{j=1}^N\sum_{k=1}^N x_jy_k}, \
\eeqn
\beqn\label{HIM_MIX_WF_DEN_MAT2}
& &
\Psi(x_1,\ldots,x_N,y_1,\ldots,y_N)\Psi^\ast(x'_1,\ldots,x'_N,y'_1,\ldots,y'_N) = 
\left(\frac{m\Omega}{\pi}\right)^{N-1}
\left(\frac{M_{12}\Omega_{12}}{\pi}\right)^{\frac{1}{2}}
\left(\frac{M\omega}{\pi}\right)^{\frac{1}{2}} \times \nonumber \\
& &
\times
e^{-\frac{\alpha}{2}\sum_{j=1}^N \left(x_j^2 + {x'_j}^2 +  y_j^2 + {y'_j}^2\right) -
\beta \sum_{1\le j < k}^N \left(x_jx_k+x'_jx'_k+y_jy_k+y'_jy'_k\right)
+ \gamma \sum_{j=1}^N\sum_{k=1}^N \left(x_jy_k+x'_jy'_k\right)}, \
\eeqn
with
\beqn\label{HIM_MIX_WF_DEN_MAT3}
& &
\Omega = \sqrt{\omega^2+\frac{2N}{m}\left(\lambda+\lambda_{12}\right)}, \qquad
\Omega_{12} = \sqrt{\omega^2+\frac{4N}{m}\lambda_{12}}, \nonumber \\
& &
\alpha = m\Omega + \beta, \qquad
\beta = \frac{m}{2N}\left(\Omega_{12}+\omega-2\Omega\right), \qquad
\gamma=\frac{m}{2N}\left(\Omega_{12}-\omega\right), \
\eeqn
\eeqs
and the relative center-of-mass $M_{12}=\frac{m}{2N}$ and center-of-mass $M=2mN$ masses.
The wavefunction and similarly the many-particle density of the mixture 
depend on two dressed frequencies, $\Omega$ and $\Omega_{12}$,
and consist
of three parts:
One-body part with coefficient $\alpha$,
intra-species two-body coupling with coefficient $\beta$,
and inter-species two-body coupling with coefficient $\gamma$.
Whereas $\alpha$ and $\beta$ depend on the intra-species and inter-species interactions,
$\gamma$ depends on the inter-species interaction only.

Another issue worth mentioning is the stability of the mixture.
For a stable, i.e., bound,
mixture both dressed frequencies $\Omega$ and $\Omega_{12}$ must be positive.
This implies the conditions $\lambda+\lambda_{12} > - \frac{m\omega^2}{2N}$ and
$\lambda_{12} > - \frac{m\omega^2}{4N}$, respectively, on the interactions.
In other words, the inter-species interaction $\lambda_{12}$ is bound from below,
implying that the mutual repulsion between the two species cannot be too strong,
but is not bound from above,
meaning that the mutual attraction between the two species can be unlimitedly strong.
Furthermore,
the intra-species interaction $\lambda$ can take
any value as long as the inter-species interaction is sufficiently attractive, i.e.,
$\lambda > - \frac{m\omega^2}{2N} - \lambda_{12}$.
We shall return to the dressed
frequencies $\Omega$ and $\Omega_{12}$ below.

\subsection{Intra-species natural pair functions}\label{PAIR_2}

The intra-species reduced density matrices are defined when all bosons of the other type are integrated out.
We concentrate in what follows on the reduced one-particle
and in particular the two-particle density matrices of species $1$,
\beqn\label{RDMs_1_2}
& & \rho_1^{(1)}(x,x') = N \int dx_2 \cdots dx_N dy_1 \cdots dy_N 
\Psi(x,x_2,\ldots,x_N,y_1,\ldots,y_N) \times \nonumber \\
& &
\times \Psi^\ast(x',x_2,\ldots,x_N,y_1,\ldots,y_N), \nonumber \\
& & \rho_1^{(2)}(x_1,x_2,x'_1,x'_2) = N(N-1) \int dx_3 \cdots dx_N dy_1 \cdots dy_N 
\Psi(x_1,x_2,x_3,\ldots,x_N,y_1,\ldots,y_N) \times \nonumber \\
& &
\times \Psi^\ast(x'_1,x'_2,x_3,\ldots,x_N,y_1,\ldots,y_N). \
\eeqn
In a symmetric mixture, the corresponding reduced density matrices of species $2$, 
$\rho_2^{(1)}(y,y')$ and
$\rho_2^{(2)}(y_1,y_2,y'_1,y'_2)$,
are the same and need not be repeated.

The reduction of the many-particle density (\ref{HIM_MIX_WF_DEN_MAT2})
to its finite-order reduced density matrices is
somewhat lengthly and given in \cite{HIM_MIX_RDM}.
We start from the final expression for the intra-species reduced one-particle density matrix
which is given by
\beqn\label{1_RDM}
& & \rho_1^{(1)}(x,x') = N \left(\frac{\alpha+C_{1,0}}{\pi}\right)^{\frac{1}{2}}
e^{-\frac{\alpha}{2}\left(x^2+{x'}^2\right)} 
e^{- \frac{1}{4} C_{1,0} \left(x+x'\right)^2} = \nonumber \\
& &
= N \left(\frac{\alpha+C_{1,0}}{\pi}\right)^{\frac{1}{2}}
e^{-\frac{\alpha+\frac{C_{1,0}}{2}}{2}\left(x^2+{x'}^2\right)} 
e^{- \frac{1}{2} C_{1,0} xx'}, \nonumber \\
& &
\alpha + C_{1,0} =
(\alpha-\beta)
\frac{\left[(\alpha-\beta) + N\beta\right]^2 - N^2\gamma^2}
{\left[(\alpha-\beta) + N\beta\right]\left[(\alpha-\beta) + (N-1)\beta\right] - N(N-1)\gamma^2}.
\eeqn
The coefficient $C_{1,0}$ governs the properties of the intra-species reduced one-particle density matrix
and reminds one that all bosons of type $2$ and all but a single boson of type $1$ are integrated out.
As might be expected, $\rho_1^{(1)}(x,x')$
depends on the three parts of the many-boson wavefunction,
i.e., on the $\alpha$, $\beta$, and $\gamma$ terms (\ref{HIM_MIX_WF_DEN_MAT1}).
In the absence of inter-species interaction, i.e., for $\gamma=0$,
the coefficient $C_{1,0}$
boils down to that of the single-species harmonic-interaction model,
see appendix \ref{APP} for further discussion.

Just as for the case of single-species bosons \cite{HIM_JCP,HIM_SCH},
the intra-species reduced one-particle density matrix (\ref{1_RDM}) can be diagonalized using Mehler's formula.
Mehler's formula can be written as follows
\beqn\label{MEHLER}
& & \left[\frac{(1-\rho)s}{(1+\rho)\pi}\right]^{\frac{1}{2}}
e^{-\frac{1}{2}\frac{(1+\rho^2)s}{1-\rho^2}\left(x^2+{x'}^2\right)} \,
e^{+\frac{2\rho s}{1-\rho^2}xx'}
= \nonumber \\
& & 
\quad = \sum_{n=0}^\infty (1-\rho)\rho^n \frac{1}{\sqrt{2^n n!}} \left(\frac{s}{\pi}\right)^{\frac{1}{4}} H_n(\sqrt{s}x) e^{-\frac{1}{2}s x^2}
\frac{1}{\sqrt{2^n n!}} \left(\frac{s}{\pi}\right)^{\frac{1}{4}} H_n(\sqrt{s}x') e^{-\frac{1}{2}s {x'}^2}, \
\eeqn
with $s>0$ and, generally, for intra-species and inter-species
reduced density matrices as well as later on for
Schmidt decomposition of the wavefunction,
$1 > \rho \ge 0$.
$H_n$ are the Hermite polynomials.

Comparing the structure of the intra-species reduced
one-particle density matrix $\rho_1^{(1)}(x,x')$
with that of Mehler's formula one readily has
\beqn\label{S_1_1}
& & s_1^{(1)} = \sqrt{\alpha\left(\alpha+C_{1,0}\right)} =
\sqrt{\frac{\alpha(\alpha-\beta)\left\{\left[(\alpha-\beta) + N\beta\right]^2 - N^2\gamma^2\right\}}
{\left[(\alpha-\beta) + N\beta\right]\left[(\alpha-\beta) + (N-1)\beta\right] - N(N-1)\gamma^2}},
\nonumber \\ \nonumber \\
& & \rho_1^{(1)} = \frac{\alpha - s_1^{(1)}}{\alpha + s_1^{(1)}} =
\frac{\sqrt{\frac{\alpha\left\{\left[(\alpha-\beta) + N\beta\right]\left[(\alpha-\beta) + (N-1)\beta\right] - N(N-1)\gamma^2\right\}}{(\alpha-\beta)\left\{\left[(\alpha-\beta) + N\beta\right]^2 - N^2\gamma^2\right\}}}-1}
{\sqrt{\frac{\alpha\left\{\left[(\alpha-\beta) + N\beta\right]\left[(\alpha-\beta) + (N-1)\beta\right] - N(N-1)\gamma^2\right\}}{(\alpha-\beta)\left\{\left[(\alpha-\beta) + N\beta\right]^2 - N^2\gamma^2\right\}}}+1},
\nonumber \\
& & 1 - \rho_1^{(1)} = \frac{2s_1^{(1)}}{\alpha + s_1^{(1)}}. \
\eeqn
Here, $1 - \rho_1^{(1)}$ is the condensate fraction of species $1$ (and of species $2$),
i.e., the fraction of condensed bosons,
and $\rho_1^{(1)}$ is the depleted fraction,
namely, the fraction of bosons residing outside the lowest, condensed mode.
$s_1^{(1)}$ is the scaling, or effective frequency, of the intra-species
natural orbitals.
The condensate fraction, depleted fraction, and scaling of the natural orbitals
are all given in closed form as a function of the number of bosons $N$, and the intra-species $\lambda$
and inter-species $\lambda_{12}$ interaction strengths.
A specific application of the general expressions
(\ref{S_1_1}) for the mixture appears below.

For the intra-species two-particle reduced density matrix we have:
\beqn\label{2_RDM}
& &
\rho_1^{(2)}(x_1,x_2,x'_1,x'_2) = 
N(N-1) \left(\frac{\alpha+C_{1,0}}{\pi}\right)^{\frac{1}{2}} \left(\frac{\alpha+C_{2,0}}{\pi}\right)^{\frac{1}{2}} \times \nonumber \\
& & 
\times e^{-\frac{\alpha}{2}\left(x_1^2+x_2^2+{x'_1}^2+{x'_2}^2\right)} e^{-\beta\left(x_1x_2 + x'_1x'_2\right)}
e^{-\frac{1}{4} C_{2,0} \left(x_1+x_2+x'_1+x'_2\right)^2}, \nonumber \\
& &
\alpha+\beta+2C_{2,0} =
(\alpha-\beta)
\frac{\left[(\alpha-\beta) + N\beta\right]^2 - N^2\gamma^2}
{\left[(\alpha-\beta) + N\beta\right]\left[(\alpha-\beta) + (N-2)\beta\right] - N(N-2)\gamma^2}, \
\eeqn
where $C_{1,0}$ is the coefficient of the intra-species reduced one-particle density matrix (\ref{1_RDM})
and the combination $\left(\alpha+\beta+2C_{2,0}\right)$ would appear shortly after.

To obtain the natural geminals of $\rho_1^{(2)}(x_1,x_2,x'_1,x'_2)$ we 
define the variables $q_1 = \frac{1}{\sqrt{2}}\left(x_1+x_2\right)$, $q_2 = \frac{1}{\sqrt{2}}\left(x_1-x_2\right)$
and $q'_1 = \frac{1}{\sqrt{2}}\left(x'_1+x'_2\right)$, $q'_2 = \frac{1}{\sqrt{2}}\left(x'_1-x'_2\right)$,
i.e., the center-of-mass and relative coordinate of
two identical bosons.
With this rotation of coordinates we have for the different terms in (\ref{2_RDM}): 
\beqn\label{TRAS1}
& & x_1^2+x_2^2+{x'_1}^2+{x'_2}^2 = q_1^2 + {q'_1}^2 + q_2^2 + {q'_2}^2, \nonumber \\
& & x_1x_2+x'_1x'_2 = \frac{1}{2}\left(q_1^2 + {q'_1}^2 - q_2^2 - {q'_2}^2\right), \nonumber \\
& & \left(x_1+x_2+x'_1+x'_2\right)^2 = 2 \left(q_1^2 + {q'_1}^2 + 2q_1q'_1\right). \
\eeqn 
Consequently, one readily finds the diagonal form
\beqn\label{2_RDM_DIAG}
& &
\rho_1^{(2)}(q_1,q'_1,q_2,q'_2) = 
N(N-1)
\left(\frac{\alpha-\beta}{\pi}\right)^{\frac{1}{2}}
e^{-\frac{\alpha-\beta}{2}\left(q_2^2 + {q'_2}^2\right)} \times \nonumber \\
& & \times
\left(\frac{\alpha+\beta+2C_{2,0}}{\pi}\right)^{\frac{1}{2}}
e^{-\frac{\alpha+\beta+C_{2,0}}{2}\left(q_1^2 + {q'_1}^2\right)}
e^{-C_{2,0} q_1q'_1},
\eeqn
where
the normalization coefficients before and after diagonalization are, of course, equal and satisfy
$\left(\alpha+C_{1,0}\right)\left(\alpha+C_{2,0}\right)=\left(\alpha-\beta\right)\left(\alpha+\beta+2C_{2,0}\right)$.

After the transformation (\ref{TRAS1}),
the first term of $\rho_1^{(2)}(q_1,q'_1,q_2,q'_2)$ is separable as a function of $q_2$ and $q'_2$ whereas,
using Mehler's formula onto the variables $q_1$ and $q'_1$,
the second term can be diagonalized. 
Thus, comparing the second term in (\ref{2_RDM_DIAG}) and Eq.~(\ref{MEHLER}) we find
\beqn\label{S_1_2}
& & s_1^{(2)} = \sqrt{\left(\alpha+\beta\right)\left(\alpha+\beta+2C_{2,0}\right)} =
\sqrt{\frac{(\alpha^2-\beta^2)\left\{\left[(\alpha-\beta) + N\beta\right]^2 - N^2\gamma^2\right\}}
{\left[(\alpha-\beta) + N\beta\right]\left[(\alpha-\beta) + (N-2)\beta\right] - N(N-2)\gamma^2}},
\nonumber \\ \nonumber \\
& & \rho_1^{(2)} = \frac{\left(\alpha + \beta\right) - s_1^{(2)}}{\left(\alpha +\beta\right) + s_1^{(2)}} =
\frac{\sqrt{\frac{\left(\alpha+\beta\right)\left\{\left[(\alpha-\beta) + N\beta\right]\left[(\alpha-\beta) + (N-2)\beta\right] - N(N-2)\gamma^2\right\}}{(\alpha-\beta)\left\{\left[(\alpha-\beta) + N\beta\right]^2 - N^2\gamma^2\right\}}}-1}
{\sqrt{\frac{\left(\alpha+\beta\right)\left\{\left[(\alpha-\beta) + N\beta\right]\left[(\alpha-\beta) + (N-2)\beta\right] - N(N-2)\gamma^2\right\}}{(\alpha-\beta)\left\{\left[(\alpha-\beta) + N\beta\right]^2 - N^2\gamma^2\right\}}}+1},
\nonumber \\
& & 1 - \rho_1^{(2)} = \frac{2s_1^{(2)}}{\alpha + s_1^{(2)}}. \
\eeqn
With expressions (\ref{S_1_2}),
the decomposition of the intra-species reduced two-particle density matrix in terms of its natural
geminals is explicitly given by
\beqn\label{2_RDM_NAT_GEM}
& &
\rho_1^{(2)}(x_1,x_2,x'_1,x'_2) = 
N(N-1) \sum_{n=0}^\infty \left(1-\rho_1^{(2)}\right)\left(\rho_1^{(2)}\right)^n
\Phi^{(2)}_{1,n}(x_1,x_2) \Phi^{(2),\ast}_{1,n}(x'_1,x'_2), \nonumber \\
& & \Phi^{(2)}_{1,n}(x_1,x_2) =
\frac{1}{\sqrt{2^n n!}}
\left(\frac{s_1^{(2)}}{\pi}\right)^{\frac{1}{4}}
H_n\left(\sqrt{\frac{s_1^{(2)}}{2}}\left(x_1+x_2\right)\right)
e^{-\frac{1}{4}s_1^{(2)}\left(x_1+x_2\right)^2} \times \nonumber \\
& & \times
\left(\frac{\alpha-\beta}{\pi}\right)^{\frac{1}{4}}
e^{-\frac{1}{4}\left(\alpha-\beta\right)\left(x_1-x_2\right)^2}.\
\eeqn
Equation (\ref{2_RDM_NAT_GEM}) is a general result on
the inter-species natural geminals of the mixture.
Together with (\ref{S_1_2}) they imply
that
$1 - \rho_2^{(1)}$ is the fraction of condensed pairs of species $1$ (and of species $2$),
$\rho_1^{(2)}$ is the fraction of depleted pairs,
i.e., the fraction of pairs residing outside the lowest, condensed natural geminal,
and 
$s_1^{(2)}$ is the scaling, or effective frequency, of the intra-species
natural pair functions.
The intra-species natural geminals along with their condensate and depleted fractions
are prescribed as explicit functions of the number of bosons $N$, and the intra-species $\lambda$
and inter-species $\lambda_{12}$ interactions.
A specific application of the general decomposition
(\ref{S_1_2}),
(\ref{2_RDM_NAT_GEM})
to natural geminals of the mixture is provided below.
Finally, we point out that 
the generalization to higher-order intra-species reduced density matrices and corresponding natural
functions follows the above pattern
and will not be discussed further here.

Let us work out an explicit application
where we shall find and analyze fragmentation of identical bosons' pairs.
Consider the specific scenario where $\lambda+\lambda_{12}=0$,
i.e., that the intra-species interaction is inverse to and `compensates'
the effect of the inter-species interaction on each of the species
in the manner that the intra-species frequency is that of non-interacting particles,
$\Omega=m\omega$.
Then, the coefficients of the three parts of the wavefunction simplify and one has
$\alpha=m\omega+\beta = m\omega \left[1 + \frac{1}{2N}\left(\frac{\Omega_{12}}{\omega}-1\right)\right]$
and
$\beta=\gamma=\frac{m}{2N}\left(\Omega_{12}-\omega\right)$.
Consequently, expressions (\ref{S_1_1}) and (\ref{S_1_2}) simplify and the intra-species reduced
one-particle and two-particle density matrices can be evaluated further.
Thus we readily find
\beqs\label{S_INTRA_EXAM}
\beqn\label{S_1_1_EXAM}
& & s_1^{(1)} = m\omega \sqrt{\frac{1 + \frac{1}{2N}\left(\frac{\Omega_{12}}{\omega}-1\right)}
{1 + \frac{1}{2N}\left(\frac{\omega}{\Omega_{12}}-1\right)}},
\nonumber \\ \nonumber \\
& & \rho_1^{(1)} = \frac{\sqrt{\left[1 + \frac{1}{2N}\left(\frac{\Omega_{12}}{\omega}-1\right)\right]
\left[1 + \frac{1}{2N}\left(\frac{\omega}{\Omega_{12}}-1\right)\right]}-1}
{\sqrt{\left[1 + \frac{1}{2N}\left(\frac{\Omega_{12}}{\omega}-1\right)\right]
\left[1 + \frac{1}{2N}\left(\frac{\omega}{\Omega_{12}}-1\right)\right]}+1},
\nonumber \\
& & 1 - \rho_1^{(1)} = \frac{2}
{\sqrt{\left[1 + \frac{1}{2N}\left(\frac{\Omega_{12}}{\omega}-1\right)\right]
\left[1 + \frac{1}{2N}\left(\frac{\omega}{\Omega_{12}}-1\right)\right]}+1}\
\eeqn
for the intra-species reduced one-particle density matrix,
where $\alpha+C_{1,0} = m\omega \frac{1}{1 + \frac{1}{2N}\left(\frac{\omega}{\Omega_{12}}-1\right)}$ is used,
and
\beqn\label{S_1_2_EXAM}
& & s_1^{(2)} = m\omega \sqrt{\frac{1 + \frac{1}{N}\left(\frac{\Omega_{12}}{\omega}-1\right)}
{1 + \frac{1}{N}\left(\frac{\omega}{\Omega_{12}}-1\right)}},
\nonumber \\ \nonumber \\
& & \rho_1^{(2)} = \frac{\sqrt{\left[1 + \frac{1}{N}\left(\frac{\Omega_{12}}{\omega}-1\right)\right]
\left[1 + \frac{1}{N}\left(\frac{\omega}{\Omega_{12}}-1\right)\right]}-1}
{\sqrt{\left[1 + \frac{1}{N}\left(\frac{\Omega_{12}}{\omega}-1\right)\right]
\left[1 + \frac{1}{N}\left(\frac{\omega}{\Omega_{12}}-1\right)\right]}+1},
\nonumber \\
& & 1 - \rho_1^{(2)} = \frac{2}
{\sqrt{\left[1 + \frac{1}{N}\left(\frac{\Omega_{12}}{\omega}-1\right)\right]
\left[1 + \frac{1}{N}\left(\frac{\omega}{\Omega_{12}}-1\right)\right]}+1} \
\eeqn
\eeqs
for the intra-species reduced two-particle density matrix,
where 
$\alpha+\beta = m\omega \left[1 + \frac{1}{N}\left(\frac{\Omega_{12}}{\omega}-1\right)\right]$ and
$\alpha+\beta+2C_{2,0} = m\omega \frac{1}{1 + \frac{1}{N}\left(\frac{\omega}{\Omega_{12}}-1\right)}$ are utilized.
We see that fragmentation of identical pairs and bosons
is governed by the ratio $\frac{\Omega_{12}}{\omega}$ and its inverse $\frac{\omega}{\Omega_{12}}$,
meaning that it takes place both at the attractive and repulsive sectors of interactions.
Moreover, the condensed and depleted fractions
of the pairs and bosons are symmetric to interchanging $\frac{\Omega_{12}}{\omega}$ and 
$\frac{\omega}{\Omega_{12}}$, see discussion below.

Let us analyze explicitly macroscopic fragmentation of geminals, i.e.,
when there is macroscopic occupation of more than a single
intra-species natural pair function of\break\hfill $\rho_1^{(2)}(x_1,x_2,x'_1,x'_2)$.
As a reference, we also refer to the corresponding and standardly defined
macroscopic fragmentation of the intra-species natural orbitals of $\rho_1^{(1)}(x,x')$. 
The structure of the eigenvalues,
emanating from Mehler's formula and its applicability to the various reduced density matrices,
suggests that, say, the `middle' value $\rho=1-\rho=\frac{1}{2}$, i.e., when the condensed and depleted fractions are equal,
is a convenient manifestation of macroscopic fragmentation.
Indeed, for this value the first few natural occupation fractions $(1-\rho)\rho^n$, $n=0,1,2,3,4,\ldots$ are
\beq\label{rho_half}
\frac{1}{2}, \quad \frac{1}{4}, \quad \frac{1}{8}, \quad \frac{1}{16}, \quad \frac{1}{32}, \ldots,
\eeq
namely,
there is $50\%$ occupation of the first natural geminal,
$25\%$ occupation of the second,
$12.5\%$ of the third,
$6.25\%$ of the fourth,
$3.125\%$ of the fifth,
and so on.
For brevity,
we refer to the fragmentation values in (\ref{rho_half}) as $50\%$ fragmentation. 

Now, one can compute for which ratio $\frac{\Omega_{12}}{\omega}$,
or, equivalently, for which inter-species interaction
$\lambda_{12} = \frac{m\omega^2}{4N}\left[\left(\frac{\Omega_{12}}{\omega}\right)^2-1\right]$,
the intra-species reduced
two-particle and one-particle density matrices are macroscopically fragmented as in (\ref{rho_half}).
Thus, solving (\ref{S_1_1_EXAM}) for $50\%$ natural-orbital fragmentation we find
\beqn\label{FRAG_INTRA_1}
 \rho_1^{(1)}=\frac{1}{2} \quad
\Longrightarrow \quad
\frac{\Omega_{12}}{\omega} =
\left(1 + \frac{8N^2}{N-\frac{1}{2}}\right) \pm \sqrt{\left(1 + \frac{8N^2}{N-\frac{1}{2}}\right)^2 - 1},
\eeqn
and working out (\ref{S_1_2_EXAM}) for $50\%$ natural-geminal fragmentation we get
\beqn\label{FRAG_INTRA_2}
 \rho_1^{(2)}=\frac{1}{2} \quad
\Longrightarrow \quad
\frac{\Omega_{12}}{\omega} = 
\sqrt{1+\frac{4N}{m\omega^2}\lambda_{12}} = 
\left(1 + \frac{4N^2}{N-1}\right) \pm \sqrt{\left(1 + \frac{4N^2}{N-1}\right)^2 - 1}.
\eeqn
There are two `reciprocate' solutions for both
the natural geminals and natural orbitals:
We see that $50\%$ fragmentation occurs for strong attractions,
i.e., when $\frac{\Omega_{12}}{\omega}$ is large,
or near the border of stability for repulsions, that is when $\frac{\Omega_{12}}{\omega}$ is close to zero.
Also, to achieve the same degree of $50\%$ with a larger number $N$ of species $1$ (and species $2$) bosons,
a stronger attraction or repulsion is needed.
Finally, comparing natural-geminal with natural-orbital fragmentation at the same $50\%$ value,
one sees from (\ref{FRAG_INTRA_2}) and (\ref{FRAG_INTRA_1}) that slightly weaker
interactions, attractions or repulsions, are needed for the former.

It is also useful to register the one-particle and two-particle densities, i.e.,
the diagonal parts $\rho_1^{(1)}(x)=\rho_1^{(1)}(x,x'=x)$ and $\rho_1^{(2)}(x_1,x_2)=\rho_1^{(2)}(x_1,x_2,x'_1=x_1,x'_2=x_2)$,
which read
\beqn\label{DENSITIES_1_2_EXAM}
& & \rho_1^{(1)}(x) =
N \left(\frac{\alpha+C_{1,0}}{\pi}\right)^{\frac{1}{2}} e^{-\left(\alpha+C_{1,0}\right)x^2} = 
N \left(\frac{m\omega}{\pi\left[1 + \frac{1}{2N}\left(\frac{\omega}{\Omega_{12}}-1\right)\right]}\right)^{\frac{1}{2}}
e^{-\frac{m\omega}{1 + \frac{1}{2N}\left(\frac{\omega}{\Omega_{12}}-1\right)}x^2},
\nonumber \\
& & \rho_1^{(2)}(x_1,x_2) = 
N(N-1)
\left(\frac{\alpha-\beta}{\pi}\right)^{\frac{1}{2}}
e^{-\frac{\alpha-\beta}{2}\left(x_1-x_2\right)^2}
\left(\frac{\alpha+\beta+2C_{2,0}}{\pi}\right)^{\frac{1}{2}}
e^{-\frac{\alpha+\beta+2C_{2,0}}{2}\left(x_1+x_2\right)^2} = \nonumber \\
& &
= N(N-1)
\left(\frac{m\omega}{\pi}\right)^{\frac{1}{2}}
e^{-\frac{m\omega}{2}\left(x_1-x_2\right)^2}
\left(\frac{m\omega}{\pi\left[1 + \frac{1}{N}\left(\frac{\omega}{\Omega_{12}}-1\right)\right]}\right)^{\frac{1}{2}}
e^{-\frac{m\omega}{2\left[1 + \frac{1}{N}\left(\frac{\omega}{\Omega_{12}}-1\right)\right]}\left(x_1+x_2\right)^2}.
\eeqn
From the densities (\ref{DENSITIES_1_2_EXAM}) we can infer a measure for the size
of identical pairs' and bosons' clouds using the widths of the respective Gaussian functions therein.
Thus, we have
\beqs\label{WIDTH_INTRA}
\beqn\label{WIDTH_INTRA_GEN}
& &
\sigma_{1,x}^{(1)} = \sqrt{\frac{1 + \frac{1}{2N}\left(\frac{\omega}{\Omega_{12}}-1\right)}{2m\omega}}, \nonumber \\
& &
\sigma_{1,\frac{x_1+x_2}{\sqrt{2}}}^{(2)} = \sqrt{\frac{1 + \frac{1}{N}\left(\frac{\omega}{\Omega_{12}}-1\right)}{2m\omega}},
\quad
\sigma_{1,\frac{x_1-x_2}{\sqrt{2}}}^{(2)} = \sqrt{\frac{1}{2m\omega}}. \
\eeqn
To assess the combined impact of the intra-species and inter-species interactions
atop the fragmentation of the reduced density matrices,
it is useful to compute the sizes (\ref{WIDTH_INTRA_GEN}) for large inter-species attractions or 
inter-species repulsions at the border of stability.
One finds, respectively,
\beqn\label{WIDTH_INTRA_LIM}
& & \lim_{\frac{\Omega_{12}}{\omega} \to \infty} \sigma_{1,x}^{(1)} = 
\sqrt{\frac{1 - \frac{1}{2N}}{2m\omega}}, \qquad
\sigma_{1,x}^{(1)} \longrightarrow \infty \quad \mathrm{for} \quad \frac{\Omega_{12}}{\omega} \to 0^+, \nonumber \\
& & \lim_{\frac{\Omega_{12}}{\omega} \to \infty} \sigma_{1,\frac{x_1+x_2}{\sqrt{2}}}^{(2)} = 
\sqrt{\frac{1 - \frac{1}{N}}{2m\omega}}, \qquad
\sigma_{1,\frac{x_1+x_2}{\sqrt{2}}}^{(2)} \longrightarrow \infty \quad \mathrm{for} \quad \frac{\Omega_{12}}{\omega} \to 0^+,
\
\eeqn
\eeqs
where $\sigma_{1,\frac{x_1-x_2}{\sqrt{2}}}^{(2)}$ is independent of the interactions.
Interestingly, the size of the densities 
for strong inter-species attractions, which is accompanied by strong intra-species repulsions because $\lambda+\lambda_{12}=0$,
saturates at about the trap's size and does not depend on the strengths of interactions.
In other words, a high degree of fragmentation is possible in the mixture without shrinking of the density due to strong inter-species
attractive interaction or expansion of the intra-species densities due to strong intra-species repulsive interaction. 
For the sake of comparative analysis, it is instructive to make contact with
fragmentation of single-species bosons in the harmonic-interaction model,
see appendix \ref{APP}.

\subsection{Inter-species natural pair functions}\label{PAIR_3}

As mentioned above,
in a mixture of two types of identical bosons there are other kinds of pairs, namely,
pairs of distinguishable particles.
If we are to examine the lowest-order inter-species reduced density matrix,
we can ask regarding distinguishable pairs questions analogous to those asked concerning identical pairs.
The purpose of this subsection is to
derive the relevant tools and answer such questions.

The inter-species reduced two-particle density matrix,
i.e., the lowest-oder inter-species quantity, is defined from the all-particle density as
\beqn\label{2_RDM_12_SYM_GEN}
& & \rho_{12}^{(2)}(x,x',y,y') = N^2 \int dx_2 \cdots dx_N dy_2 \cdots dy_N 
\Psi(x,x_2,\ldots,x_N,y,y_2,\ldots,y_N) \times \nonumber \\
& &
\times \Psi^\ast(x',x_2,\ldots,x_N,y',y_2,\ldots,y_N). \
\eeqn
For the harmonic-interaction model of the symmetric mixture it
can be computed analytically and, starting from (\ref{HIM_MIX_WF_DEN_MAT2}),
is given by \cite{HIM_MIX_RDM}
\beqs\label{2_RDM_12_SYM}
\beqn\label{2_RDM_12_SYM_DEN}
& & \rho_{12}^{(2)}(x,x',y,y') = N^2
\left[\frac{(\alpha_1+C_{1,1})^2-D_{1,1}^2}{\pi^2}\right]^\frac{1}{2} \times \nonumber \\
& & \times e^{-\frac{\alpha_1}{2} \left(x^2+{x'}^2+y^2+{y'}^2\right)}
e^{-\frac{1}{4}C_{1,1}\left[\left(x+x'\right)^2+\left(y+y'\right)^2\right]}
e^{+\frac{1}{2}D_{1,1} \left(x+x'\right)\left(y+y'\right)} 
e^{+\frac{1}{2}D'_{1,1} \left(x-x'\right)\left(y-y'\right)}, \
\eeqn
where
\beqn\label{2_RDM_12_SYM_COEFF}
& & \alpha + C_{1,1} \mp D_{1,1} =
(\alpha-\beta)
\frac{(\alpha-\beta) + N(\beta\mp \gamma)}
{(\alpha-\beta) + (N-1)(\beta\mp \gamma)},
\nonumber \\
& & D'_{1,1} = \gamma.
\eeqn
\eeqs
We see that the structure of the inter-species reduced two-particle density matrix
is more involved than that of the intra-species reduced two-particle density matrix
as well as that of the product of the two, species $1$ and species $2$ intra-species reduced one-particle density matrices.
Nonetheless, it can be diagonalized.

To diagonalize
$\rho_{12}^{(2)}(x,x',y,y')$
one must couple and make linear combinations of
coordinates associated with distinguishable bosons.
Defining $u = \frac{1}{\sqrt{2}}\left(x+y\right)$, $v = \frac{1}{\sqrt{2}}\left(x-y\right)$
and 
$u'= \frac{1}{\sqrt{2}}\left(x'+y'\right)$, $v'= \frac{1}{\sqrt{2}}\left(x'-y'\right)$
we have for the different terms in (\ref{2_RDM_12_SYM}):
\beqn\label{TRAS3}
& & x^2+y^2+{x'}^2+{y'}^2 = u^2 + {u'}^2 + v^2 + {v'}^2, \\
& & \left(x+x'\right)^2+\left(y+y'\right)^2 = \left(u+u'\right)^2 + \left(v+v'\right)^2 =
u^2 + {u'}^2 + v^2 + {v'}^2 + 2\left(uu'+vv'\right), \nonumber \\
& & \left(x \pm x'\right)\left(y \pm y'\right) = \frac{1}{2}\left[\left(u\pm u'\right)^2 - \left(v\pm v'\right)^2\right] =
\frac{1}{2}\left(u^2 + {u'}^2 - v^2 - {v'}^2\right) \pm \left(uu'-vv'\right). \nonumber \
\eeqn
Consequently, we readily find the decomposition
\beqn\label{2_RDM_12_SYM_DIAG}
& &
\rho_{12}^{(2)}(u,u',v,v') = N^2
\left(\frac{\alpha_1+C_{1,1}-D_{1,1}}{\pi}\right)^\frac{1}{2}
e^{-\frac{\alpha_1+\frac{C_{1,1}}{2}-\frac{D_{1,1}+D'_{1,1}}{2}}{2}\left(u^2 + {u'}^2\right)}
e^{-\frac{1}{2}\left[C_{1,1}-\left(D_{1,1}-D'_{1,1}\right)\right]uu'} \times \nonumber \\
& & \times 
\left(\frac{\alpha_1+C_{1,1}+D_{1,1}}{\pi}\right)^\frac{1}{2}
e^{-\frac{\alpha_1+\frac{C_{1,1}}{2}+\frac{D_{1,1}+D'_{1,1}}{2}}{2}\left(v^2 + {v'}^2\right)}
e^{-\frac{1}{2}\left[C_{1,1}+\left(D_{1,1}-D'_{1,1}\right)\right]vv'},
\eeqn
where the normalizations after and before diagonalization are, of course, equal.
As might be expected,
since the structure of $\rho_{12}^{(2)}(x,x',y,y')$ is more involved than that
of $\rho_1^{(2)}(x_1,x_2,x'_1,x'_2)$ the diagonalization of the former is more intricate.
Fortunately, we can do that using the application of Mehler's formula twice,
on the appropriately-constructed inter-species `mixed' coordinates $u, u'$ and $v, v'$.
We thus get
\beqn\label{S_12_2}
& & s_{12,\pm}^{(2)} = \sqrt{\left(\alpha \mp D'_{1,1}\right)\left(\alpha+C_{1,1} \mp D_{1,1}\right)} =
\sqrt{(\alpha\mp \gamma)(\alpha-\beta)
\frac{(\alpha-\beta) + N(\beta\mp \gamma)}
{(\alpha-\beta) + (N-1)(\beta\mp \gamma)}},
\nonumber \\
& & \rho_{12,\pm}^{(2)} = \frac{\left(\alpha \mp D'_{1,1}\right) - s_{12,\pm}^{(2)}}{\left(\alpha \mp D'_{1,1}\right) + s_{12,\pm}^{(2)}} =
\frac{\frac{(\alpha\mp \gamma)\left[(\alpha-\beta) + (N-1)(\beta\mp \gamma)\right]}{(\alpha-\beta)\left[(\alpha-\beta) + N(\beta\mp \gamma)\right]}-1}{\frac{(\alpha\mp \gamma)\left[(\alpha-\beta) + (N-1)(\beta\mp \gamma)\right]}{(\alpha-\beta)\left[(\alpha-\beta) + N(\beta\mp \gamma)\right]}+1},
\nonumber \\
& & 1 - \rho_{12,\pm}^{(2)} = \frac{2s_{12,\pm}^{(2)}}{\left(\alpha \mp D'_{1,1}\right) + s_{12,\pm}^{(2)}}, \
\eeqn
where the ``$+$'' terms quantify the fragmentation in the $u,u'$ part of the intra-species reduced two-particle density matrix and
the ``$-$'' terms quantify the fragmentation in the $v,v'$ part of the intra-species reduced two-particle density matrix,
also see below.
Equation (\ref{S_12_2}) is one of the main results of the present work and bears
a clear and appealing physical meaning,
that pairs made of distinguishable bosons can be fragmented,
and that this fragmentation is governed
by the center-of-mass and 
by a relative coordinate of distinguishable bosons.
We shall return to this point in what follows.

We can now prescribe the decomposition of the
inter-species reduced two-particle density matrix
to its distinguishable natural pair functions which is given by
\beqn\label{12_2_RDM_NAT_DIS_GEM}
& &
\rho_{12}^{(2)}(x,x',y,y') = \nonumber \\
& & = N^2 
\sum_{n_+=0}^\infty \sum_{n_-=0}^\infty
\left(1-\rho_{12,+}^{(2)}\right)
\left(1-\rho_{12,-}^{(2)}\right)
\left(\rho_{12,+}^{(2)}\right)^{\!n_+}
\left(\rho_{12,-}^{(2)}\right)^{\!n_-}
\Phi^{(2)}_{12,n_+,n_-}(x,y) \Phi^{(2),\ast}_{12,n_+,n_-}(x',y'), \nonumber \\
& & \Phi^{(2)}_{12,n_+,n_-}(x,y) =
\frac{1}{\sqrt{2^{n_+} {n_+}!}}
\left(\frac{s_{12,+}^{(2)}}{\pi}\right)^{\frac{1}{4}}
H_{n_+}\left(\sqrt{\frac{s_{12,+}^{(2)}}{2}}\left(x+y\right)\right)
e^{-\frac{1}{4}s_{12,+}^{(2)}\left(x+y\right)^2} \times \nonumber \\
& & \times
\frac{1}{\sqrt{2^{n_-} {n_-}!}}
\left(\frac{s_{12,-}^{(2)}}{\pi}\right)^{\frac{1}{4}}
H_{n_-}\left(\sqrt{\frac{s_{12,-}^{(2)}}{2}}\left(x-y\right)\right)
e^{-\frac{1}{4}s_{12,-}^{(2)}\left(x-y\right)^2}.\
\eeqn
All in all,
(\ref{12_2_RDM_NAT_DIS_GEM}) implies
that the distinguishable-pair `condensed fraction' is given by\break\hfill
$\left(1-\rho_{12,+}^{(2)}\right)\left(1-\rho_{12,-}^{(2)}\right)$ 
and the respective depleted fraction by
$1-\left(1-\rho_{12,+}^{(2)}\right)\left(1-\rho_{12,-}^{(2)}\right)=
\rho_{12,+}^{(2)}+\rho_{12,-}^{(2)}-\rho_{12,+}^{(2)}\rho_{12,-}^{(2)}$. 
Each of the inter-species `mixed' coordinates $\frac{x\pm y}{\sqrt{2}}$
carry its own scaling, $s_{12,\pm}^{(2)}$.
The distinguishable natural geminals 
$ \Phi^{(2)}_{12,n_+,n_-}(x,y)$
are, needless to say, orthonormal to each other. 

We proceed now for an application.
We considered above the specific case of $\lambda+\lambda_{12}=0$
which leads to 
$\Omega=m\omega$,
$\alpha=m\omega \left[1 + \frac{1}{2N}\left(\frac{\Omega_{12}}{\omega}-1\right)\right]$, and
$\beta=\gamma=\frac{m}{2N}\left(\Omega_{12}-\omega\right)$.
To evaluate $\rho_{12}^{(2)}(x,x',y,y')$
we also need the combinations
$\left(\alpha + C_{1,1} - D_{1,1}\right) = m \omega$
and
$\left(\alpha - D'_{1,1}\right) = m\omega$ for the ``$+$'' branch
as well as
$\left(\alpha + C_{1,1} + D_{1,1}\right) =
m\omega \frac{1}{1 + \frac{1}{N}\left(\frac{\omega}{\Omega_{12}}-1\right)}$
and
$\left(\alpha + D'_{1,1}\right) =
m\omega \left[1 + \frac{1}{N}\left(\frac{\Omega_{12}}{\omega}-1\right)\right]$
for the ``$-$'' branch.

Thus, expressions (\ref{S_12_2}) can readily be evaluated and the following picture
of inter-species fragmentation is found:
\beqs\label{S2_INTER_EXAM}
\beqn\label{S2_12_+_EXAM}
& & s_{12,+}^{(2)} = m\omega,
\qquad
\rho_{12,+}^{(2)} = 0,
\qquad
1 - \rho_{12,+}^{(2)} = 1, \
\eeqn
indicating that there is no contribution to fragmentation from the symmetric `mixed' coordinate $u, u'$.
On the other end,
\beqn\label{S2_12_-_EXAM}
& & s_{12,-}^{(2)} = m\omega \sqrt{\frac{1 + \frac{1}{N}\left(\frac{\Omega_{12}}{\omega}-1\right)}
{1 + \frac{1}{N}\left(\frac{\omega}{\Omega_{12}}-1\right)}},
\nonumber \\ \nonumber \\
& & \rho_{12,-}^{(2)} = \frac{\sqrt{\left[1 + \frac{1}{N}\left(\frac{\Omega_{12}}{\omega}-1\right)\right]
\left[1 + \frac{1}{N}\left(\frac{\omega}{\Omega_{12}}-1\right)\right]}-1}
{\sqrt{\left[1 + \frac{1}{N}\left(\frac{\Omega_{12}}{\omega}-1\right)\right]
\left[1 + \frac{1}{N}\left(\frac{\omega}{\Omega_{12}}-1\right)\right]}+1},
\nonumber \\
& & 1 - \rho_{12,-}^{(2)} = \frac{2}
{\sqrt{\left[1 + \frac{1}{N}\left(\frac{\Omega_{12}}{\omega}-1\right)\right]
\left[1 + \frac{1}{N}\left(\frac{\omega}{\Omega_{12}}-1\right)\right]}+1}, \
\eeqn
\eeqs
namely, that the fragmentation fully originates from the asymmetric `mixed' coordinate $v, v'$.
We conclude that,
whereas fragmentation of identical pairs is associated with their center-of-mass coordinate,
fragmentation of distinguishable pairs is linked, in this explicit case, only with a relative coordinate between
two distinguishable bosons. 
Interestingly,
the degree of intra-species and inter-species pair fragmentation is the same
in the specific case considered,
despite pertaining to different parts of the mixtures' many-boson wavefunction.
Furthermore,
there are different numbers of pairs:
$\frac{N}{2}$ intra-species identical pairs (for each of the species)
and $N$ inter-species pairs of distinguishable bosons.

Now, one can compute the ratio $\frac{\Omega_{12}}{\omega}=\sqrt{1+\frac{4N}{m\omega^2}\lambda_{12}}$
for which the inter-species reduced two-particle density matrix is $50\%$ fragmented as in (\ref{rho_half}).
Since $\rho_{12,+}^{(2)}=0$ does not contribute,
the only contribution to fragmentation comes from
$\rho_{12,-}^{(2)}$.
Thus, solving (\ref{S2_12_-_EXAM}) for $50\%$ distinguishable-pair-function fragmentation we obtain
\beqn\label{FRAG_INTER_2}
\rho_{12,-}^{(2)}=\frac{1}{2} \quad
\Longrightarrow \quad
\frac{\Omega_{12}}{\omega} =
\left(1 + \frac{4N^2}{N-1}\right) \pm \sqrt{\left(1 + \frac{4N^2}{N-1}\right)^2 - 1}.
\eeqn
As above, there are two `reciprocate' solutions,
one for strong inter-species attraction and the second close to the border of stability for intermediate-strength
inter-species repulsion.
We remind that since $\lambda+\lambda_{12}=0$ in our example,
the respective intra-species interaction is opposite in sign.
Also, to achieve the same degree of $50\%$ fragmentation
with a larger number $N$ of
distinguishable pairs,
a stronger inter-species attraction or repulsion is needed.
Furthermore, as discussed above,
comparing distinguishable-pair and identical-pair fragmentation at the same $50\%$ value
in this example,
one sees from (\ref{FRAG_INTER_2}) and (\ref{FRAG_INTRA_2}) that the same interaction
is needed.

Finally, we prescribe the inter-species two-particle density,
namely,
the diagonal part $\rho_{12}^{(2)}(x,y)=\rho_{12}^{(2)}(x,x'=x,y,y'=y)$,
which is given by
\beqn\label{DENSITY_2_12_EXAM}
& &
\rho_{12}^{(2)}(x,y) = N^2
\left(\frac{\alpha_1+C_{1,1}-D_{1,1}}{\pi}\right)^\frac{1}{2}
e^{-\frac{\alpha_1+C_{1,1}-D_{1,1}}{2}\left(x+y\right)^2}
\times \nonumber \\
& & \times 
\left(\frac{\alpha_1+C_{1,1}+D_{1,1}}{\pi}\right)^\frac{1}{2}
e^{-\frac{\alpha_1+C_{1,1}+D_{1,1}}{2}\left(x-y\right)^2} = \nonumber \\
& &
= N^2
\left(\frac{m\omega}{\pi}\right)^{\frac{1}{2}}
e^{-\frac{m\omega}{2}\left(x+y\right)^2}
\left(\frac{m\omega}{\pi\left[1 + \frac{1}{N}\left(\frac{\omega}{\Omega_{12}}-1\right)\right]}\right)^{\frac{1}{2}}
e^{-\frac{m\omega}{2\left[1 + \frac{1}{N}\left(\frac{\omega}{\Omega_{12}}-1\right)\right]}\left(x-y\right)^2}.
\eeqn
Next, the size of the distinguishable pairs' cloud can be assessed 
from the density (\ref{DENSITY_2_12_EXAM})
using the widths of the respective Gaussian functions.
Accordingly, we find
\beqs\label{WIDTH_INTER}
\beqn\label{WIDTH_INTER_GEN}
& & \sigma_{12,\frac{x+y}{\sqrt{2}}}^{(2)} = \sqrt{\frac{1}{2m\omega}}, \quad
\sigma_{12,\frac{x-y}{\sqrt{2}}}^{(2)} = \sqrt{\frac{1 + \frac{1}{N}\left(\frac{\omega}{\Omega_{12}}-1\right)}{2m\omega}}. \
\eeqn
To show the combined effect of the intra-species and inter-species interactions
accompanying
fragmentation of $\rho_{12}^{(2)}(x,x',y,y')$,
it is useful to compute the sizes (\ref{WIDTH_INTER_GEN}) for large inter-species attractions or 
inter-species repulsions at the border of stability.
We obtain, respectively,
\beqn\label{WIDTH_INTER_LIM}
& & \lim_{\frac{\Omega_{12}}{\omega} \to \infty} \sigma_{12,\frac{x-y}{\sqrt{2}}}^{(2)} = 
\sqrt{\frac{1 - \frac{1}{N}}{2m\omega}}, \qquad
\sigma_{12,\frac{x-y}{\sqrt{2}}}^{(2)} \longrightarrow \infty \quad \mathrm{for} \quad \frac{\Omega_{12}}{\omega} \to 0^+,
\
\eeqn
\eeqs
where $\sigma_{12,\frac{x+y}{\sqrt{2}}}^{(2)}$ is independent of the interactions.
We see that the size of the inter-species density
saturates as well at about the trap's size and does not depend on the strengths of interactions
in the limit of strong inter-species attractions.
Analogously to identical pairs,
a strong fragmentation of distinguishable pairs
is possible in the mixture without shrinking of the inter-species 
density due to strong inter-species
attractive interaction.
At the other end,
when the inter-species repulsion is close to the border of stability,
the inter-species density expands boundlessly.
Summarizing, inter-species fragmentation is governed by the ratio $\frac{\Omega_{12}}{\omega}$
and takes place both at the attractive and repulsive sectors of interactions.
For the sake of analysis, we compared the results for inter-species pair fragmentation with intra-species pair fragmentation
and discussed the similarity and differences between the respective two-particle densities
(\ref{DENSITY_2_12_EXAM}) and (\ref{DENSITIES_1_2_EXAM}).

\section{Pair of distinguishable pairs and Schmidt decomposition of the wavefunction}\label{MORE}

Following the results of the previous section
on fragmentation of distinguishable pairs,
there are two questions that warrant answers.
The first is whether inter-species fragmentation persists beyond distinguishable pairs,
say, to pairs of distinguishable pairs?
In as much as single-species and intra-species fragmentations
take place at the lowest-level reduced one-particle density matrix,
and persist at higher-level single-species reduced density matrices,
we wish to establish the result of inter-species fragmentation at the level of
higher-order reduced density matrices.
After all, the reduced two-particle density matrix is the lowest-order inter-species one.
The second question deals with the nature of the inter-species coordinates governing fragmentation.
At the level of distinguishable-pair fragmentation,
i.e., within the inter-species reduced two-particle density matrix,
one cannot unambiguously tell whether the relative center-of-mass
coordinate of the two species is involved or whether other relative inter-species coordinates govern fragmentation.
This is because for a pair of distinguishable particles one cannot
distinguish between the two types of coordinates.

As seen in the previous section,
the inter-species reduced two-particle density matrix is more intricate
than the intra-species ones, 
and consequently its diagonalization is more involved.
We derive now the inter-species reduced four-particle density matrix
and examine which `normal coordinates' govern its diagonalization.
Then, the natural four-particle functions are obtained explicitly and investigated.

Finally and as a complementary result of the techniques
used for inter-species fragmentation,
we carry the connection between inter-species and intra-species center-of-mass coordinates,
in conjunction with the usage of Mehler's formula within a mixture, further.
This is done
by constructing the Schmidt decomposition of the mixture's wavefunction
and discussing consequences of this decomposition at the limit of an infinite number of particles.

\subsection{Inter-species fragmentation in higher-order reduced density matrices}\label{MORE_FRAG}

The inter-species reduced four-particle density matrix is defined as
\beqn\label{4_RDM_12_SYM_GEN}
& & \rho_{12}^{(4)}(x_1,x_2,x'_1,x'_2,y_1,y_2,y'_1,y'_2) = N^2 \left(N-1\right)^2
\int dx_3 \cdots dx_N dy_3 \cdots dy_N \times \nonumber \\
& & \times \Psi(x_1,x_2,x_3,\ldots,x_N,y_1,y_2,y_3,\ldots,y_N)
\Psi^\ast(x'_1,x'_2,x_3,\ldots,x_N,y'_1,y'_2,y_3,\ldots,y_N). \
\eeqn
Note that here we only treat the four-particle quantity with two identical bosons per each species.
Integrating the harmonic interaction-model for symmetric mixtures we find the final expression explicitly
\beqs\label{4_RDM_12_SYM}
\beqn\label{4_RDM_12_SYM_DEN}
& & \rho_{12}^{(4)}(x_1,x_2,x'_1,x'_2,y_1,y_2,y'_1,y'_2) = N^2 \left(N-1\right)^2
\left[\frac{(\alpha+C_{1,1})^2-D_{1,1}^2}{\pi^2}\right]^\frac{1}{2}
\left[\frac{(\alpha+C_{2,2})^2-D_{2,2}^2}{\pi^2}\right]^\frac{1}{2} \times \nonumber \\
& & \times e^{-\frac{\alpha}{2} \left(x_1^2+x_2^2+{x'_1}^2+{x'_2}^2+y_1^2+y_2^2+{y'_1}^2+{y'_2}^2\right)}
e^{-\beta\left(x_1x_2+x'_1x'_2+y_1y_2+y'_1y'_2\right)} \times \nonumber \\
& & \times
e^{-\frac{1}{4}C_{2,2}\left[\left(x_1+x_2+x'_1+x'_2\right)^2+\left(y_1+y_2+y'_1+y'_2\right)^2\right]} \times \nonumber \\
& & \times e^{+\frac{1}{2}D_{2,2} \left(x_1+x_2+x'_1+x'_2\right)\left(y_1+y_2+y'_1+y'_2\right)} 
e^{+\frac{1}{2}D'_{2,2} \left(x_1+x_2-x'_1-x'_2\right)\left(y_1+y_2-y'_1-y'_2\right)}, \
\eeqn
where
\beqn\label{4_RDM_12_SYM_COEFF}
& & \alpha + \beta +2\left(C_{2,2} \mp D_{2,2}\right) =
(\alpha-\beta)
\frac{(\alpha-\beta) + N(\beta\mp \gamma)}
{(\alpha-\beta) + (N-2)(\beta\mp \gamma)},
\nonumber \\
& & D'_{2,2} = \gamma,
\eeqn
\eeqs
and $\alpha+C_{1,1}\mp D_{1,1}$ are given in (\ref{2_RDM_12_SYM_COEFF}).
The combinations of parameters $\alpha + \beta +2\left(C_{2,2} \mp D_{2,2}\right)$
would appear below shortly.

To diagonalize 
$\rho_{12}^{(4)}(x_1,x_2,x'_1,x'_2,y_1,y_2,y'_1,y'_2)$
we need to mix and rotate the coordinates of the two species
into new coordinates appropriately.
Thus, defining the new coordinates as the center-of-mass, relative center-of-mass,
and relative coordinates of two identical pairs, one for each of the species,
$u_1 = \frac{1}{2}\left[\left(x_1+x_2\right)+\left(y_1+y_2\right)\right]$,
$v_1 = \frac{1}{2}\left[\left(x_1+x_2\right)-\left(y_1+y_2\right)\right]$,
$u_2 = \frac{1}{\sqrt{2}}\left(x_1-x_2\right)$,
$v_2 = \frac{1}{\sqrt{2}}\left(y_1-y_2\right)$
and
$u'_1 = \frac{1}{2}\left[\left(x'_1+x'_2\right)\right.+\left.\left(y'_1+y'_2\right)\right]$,
$v'_1 = \frac{1}{2}\left[\left(x'_1+x'_2\right)-\left(y'_1+y'_2\right)\right]$,
$u'_2 = \frac{1}{\sqrt{2}}\left(x'_1-x'_2\right)$,
$v'_2 = \frac{1}{\sqrt{2}}\left(y'_1-y'_2\right)$,
we have for the different terms in (\ref{4_RDM_12_SYM_DEN}): 
\beqn\label{TRAS4}
& & x_1^2+x_2^2+y_1^2+y_2^2+{x'_1}^2+{x'_2}^2+{y'_1}^2+{y'_2}^2 =
u_1^2+{u'_1}^2+v_1^2+{v'_1}^2+u_2^2+{u'_2}^2+v_2^2+{v'_2}^2, \nonumber \\
& & x_1x_2+x'_1x'_2+y_1y_2+y'_1y'_2 =
\frac{1}{2}\left(u_1^2+{u'_1}^2+v_1^2+{v'_1}^2-u_2^2-{u'_2}^2-v_2^2-{v'_2}^2\right),  \nonumber \\
& & \left(x_1+x_2+x'_1+x'_2\right)^2+\left(y_1+y_2+y'_1+y'_2\right)^2 =
2\left[\left(u_1+u'_1\right)^2 + \left(v_1+v'_1\right)^2\right], \nonumber \\
& & \left[\left(x_1+x_2\right)\pm\left(x'_1+x'_2\right)\right]\left[\left(y_1+y_2\right)\pm\left(y'_1+y'_2\right)\right] = 
\left(u_1 \pm u'_1\right)^2 - \left(v_1 \pm v'_1\right)^2. \
\eeqn
Relations (\ref{TRAS4}) imply that
one could equally define inter-species linear combinations of the relative coordinates,
since
$u_2^2+v_2^2 = \left[\frac{\left(x_1-x_2\right)+\left(y_1-y_2\right)}{2}\right]^2 + 
\left[\frac{\left(x_1-x_2\right)-\left(y_1-y_2\right)}{2}\right]^2$ and
${u'_2}^2+{v'_2}^2 = \left[\frac{\left(x'_1-x'_2\right)+\left(y'_1-y'_2\right)}{2}\right]^2 + 
\left[\frac{\left(x'_1-x'_2\right)-\left(y'_1-y'_2\right)}{2}\right]^2$.
We
chose the former combinations.

Plugging (\ref{TRAS4}) into (\ref{4_RDM_12_SYM}) we readily find for the transformed
inter-species reduced four-particle density matrix
\beqn\label{4_RDM_12_SYM_DIAG}
& &
\rho_{12}^{(4)}(u_1,u'_1,v_1,v'_1,u_2,u'_2,v_2,v'_2) = N^2 \left(N-1\right)^2
\times \nonumber \\
& & \times 
\left(\frac{\alpha-\beta}{\pi}\right)^{\frac{1}{2}} e^{-\frac{\alpha-\beta}{2}\left(u_2^2 + {u'_2}^2\right)}
\left(\frac{\alpha-\beta}{\pi}\right)^{\frac{1}{2}} e^{-\frac{\alpha-\beta}{2}\left(v_2^2 + {v'_2}^2\right)} \times \nonumber \\
& & \times 
\left[\frac{\alpha+\beta+2\left(C_{2,2}-D_{2,2}\right)}{\pi}\right]^\frac{1}{2}
e^{-\frac{\alpha+\beta+C_{2,2}-\left(D_{2,2}+D'_{2,2}\right)}{2}\left(u_1^2 + {u'_1}^2\right)}
e^{-\left[C_{2,2}-\left(D_{2,2}-D'_{2,2}\right)\right]u_1u'_1} \times \nonumber \\
& & \times 
\left[\frac{\alpha+\beta+2\left(C_{2,2}+D_{2,2}\right)}{\pi}\right]^\frac{1}{2}
e^{-\frac{\alpha+\beta+C_{2,2}+\left(D_{2,2}+D'_{2,2}\right)}{2}\left(v_1^2 + {v'_1}^2\right)}
e^{-\left[C_{2,2}+\left(D_{2,2}-D'_{2,2}\right)\right]v_1v'_1}, \
\eeqn
where
the normalization coefficients before and after diagonalization are, naturally, equal and fulfill
$\left[\alpha+\left(C_{1,1}\mp D_{1,1}\right)\right]\left[\alpha+\left(C_{2,2}\mp D_{2,2}\right)\right]=
\left(\alpha-\beta\right)\left[\alpha+\beta+2\left(C_{2,2}\pm D_{2,2}\right)\right]$.

As can be seen in (\ref{4_RDM_12_SYM_DIAG}) and (\ref{2_RDM_12_SYM_DIAG}),
the similarities and differences between the structures
of $\rho_{12}^{(4)}(u_1,u'_1,v_1,v'_1,u_2,u'_2,v_2,v'_2)$ and $\rho_{12}^{(2)}(u,u',v,v')$ 
clarify the issue of which coordinates are coupled and identify the coordinates that are not.
In particular, just like for the two-particle quantity,
we can apply Mehler's formula twice,
on the appropriately-constructed inter-species `mixed coordinates' $u_1, u'_1$ and $v_1, v'_1$,
to diagonalize the inter-species reduced four-particle density matrix.
When this is done one obtains
\beqn\label{S_12_4}
& & s_{12,\pm}^{(4)} = \sqrt{\left(\alpha+\beta\mp 2D'_{2,2}\right)
\left[\alpha+\beta+2\left(C_{2,2} \mp D_{2,2}\right)\right]} = \nonumber \\
& & = \sqrt{(\alpha+\beta\mp 2\gamma)(\alpha-\beta)
\frac{(\alpha-\beta) + N(\beta\mp \gamma)}
{(\alpha-\beta) + (N-2)(\beta\mp \gamma)}},
\nonumber \\
& & \rho_{12,\pm}^{(4)} = \frac{\left(\alpha+\beta \mp 2D'_{2,2}\right) - s_{12,\pm}^{(4)}}{\left(\alpha+\beta \mp 2D'_{2,2}\right) + s_{12,\pm}^{(4)}} =
\frac{\frac{(\alpha+\beta\mp 2\gamma)\left[(\alpha-\beta) + (N-2)(\beta\mp \gamma)\right]}{(\alpha-\beta)\left[(\alpha-\beta) + N(\beta\mp \gamma)\right]}-1}
{\frac{(\alpha+\beta\mp 2\gamma)\left[(\alpha-\beta) + (N-2)(\beta\mp \gamma)\right]}{(\alpha-\beta)\left[(\alpha-\beta) + N(\beta\mp \gamma)\right]}+1},
\nonumber \\
& & 1 - \rho_{12,\pm}^{(4)} = \frac{2s_{12,\pm}^{(4)}}{\left(\alpha+\beta \mp 2D'_{2,2}\right) + s_{12,\pm}^{(4)}}, \
\eeqn
where the ``$+$'' terms quantify the fragmentation in the $u_1, u'_1$ part of the inter-species reduced four-particle density matrix and
the ``$-$'' terms determine
the fragmentation in the $v_1, v'_1$ part of the inter-species reduced four-particle density matrix.
As found and shown in (\ref{4_RDM_12_SYM_DIAG}),
there is no fragmentation due to the relative-coordinate parts $u_2, u'_2$ and $v_2, v'_2$.
Equation (\ref{S_12_4}) adds to the main results of the present work and bears
a transparent and appealing physical meaning:
In the mixture inter-species fragmentation is quantified by
the eigenvalues obtained from Mehler's formula when the latter
is applied to the mixture's center-of-mass and relative center-of-mass coordinates
of distinguishable pairs of pairs.
Extensions to larger distinguishable aggregates of species $1$ and species $2$ identical bosons in the mixture is possible along the above lines,
and will not be pursued further here.

We can now prescribe the decomposition of the intra-species reduced four-particle density matrix,
in as much as the reduced two-particle density matrix was decomposed, 
into its natural four-particle functions made of distinguishable particles which is given by
\beqn\label{12_4_RDM_NAT_DIS_GEM}
& &
\rho_{12}^{(4)}(x_1,x_2,x'_1,x'_2,y_1,y_2,y'_1,y'_2) = N^2(N-1)^2 
\sum_{n_+=0}^\infty \sum_{n_-=0}^\infty
\left(1-\rho_{12,+}^{(4)}\right)
\left(1-\rho_{12,-}^{(4)}\right) \times \nonumber \\
& & \times \left(\rho_{12,+}^{(4)}\right)^{\!n_+}
\left(\rho_{12,-}^{(4)}\right)^{\!n_-}
\Phi^{(4)}_{12,n_+,n_-}(x_1,x_2,y_1,y_2) \Phi^{(4),\ast}_{12,n_+,n_-}(x'_1,x'_2,y'_1,y'_2), \nonumber \\
& & \Phi^{(4)}_{12,n_+,n_-}(x_1,x_2,y_1,y_2) = \nonumber \\
& & =
\frac{1}{\sqrt{2^{n_+} {n_+}!}}
\left(\frac{s_{12,+}^{(4)}}{\pi}\right)^{\frac{1}{4}}
H_{n_+}\left(\frac{\sqrt{s_{12,+}^{(4)}}}{2}\left[\left(x_1+x_2\right)+\left(y_1+y_2\right)\right]\right)
e^{-\frac{1}{8}s_{12,+}^{(4)}\left[\left(x_1+x_2\right)+\left(y_1+y_2\right)\right]^2} \times \nonumber \\
& & \times
\frac{1}{\sqrt{2^{n_-} {n_-}!}}
\left(\frac{s_{12,-}^{(4)}}{\pi}\right)^{\frac{1}{4}}
H_{n_-}\left(\frac{\sqrt{s_{12,-}^{(4)}}}{2}\left[\left(x_1+x_2\right)-\left(y_1+y_2\right)\right]\right)
e^{-\frac{1}{8}s_{12,-}^{(4)}\left[\left(x_1+x_2\right)-\left(y_1+y_2\right)\right]^2} \times \nonumber \\
& & 
\times \left(\frac{\alpha-\beta}{\pi}\right)^{\frac{1}{4}}
e^{-\frac{\alpha-\beta}{4}\left(x_1-x_2\right)^2}
\left(\frac{\alpha-\beta}{\pi}\right)^{\frac{1}{4}}
e^{-\frac{\alpha-\beta}{4}\left(y_1-y_2\right)^2}. \
\eeqn
Equation (\ref{12_4_RDM_NAT_DIS_GEM}) means
that the pair-of-distinguishable-pairs `condensed fraction' is given by
the product
$\left(1-\rho_{12,+}^{(4)}\right)\left(1-\rho_{12,-}^{(4)}\right)$ 
and the respective depleted fraction is\break\hfill
$1-\left(1-\rho_{12,+}^{(4)}\right)\left(1-\rho_{12,-}^{(4)}\right)=
\rho_{12,+}^{(4)}+\rho_{12,-}^{(4)}-\rho_{12,+}^{(4)}\rho_{12,-}^{(4)}$.
The center-of-mass and relative center-of-mass coordinates of the two pairs,
$\frac{\left(x_1+x_2\right)\pm\left(y_1+y_2\right)}{2}$,
carry the respective scalings $s_{12,\pm}^{(4)}$.
The natural four-particle functions $\Phi^{(4)}_{12,n_+,n_-}(x_1,x_2,y_1,y_2)$
are enumerated by the two quantum numbers $n_+$, $n_-$ and
are obviously orthonormal to each other. 

We proceed now to examine fragmentation in this higher-order inter-species reduced density matrix.
We investigate as above the specific case of $\lambda+\lambda_{12}=0$.
To compute $\rho_{12}^{(4)}(x_1,x_2,x'_1,x'_2,y_1,y_2,y'_1,y'_2)$
we require the quantities
$\left[\alpha+\beta+2\left(C_{2,2} - D_{2,2}\right)\right] = m \omega$
and
$\left(\alpha+\beta-2D'_{2,2}\right) = m\omega$
for the ``$+$'' branch as well as
$\left[\alpha+\beta+2\left(C_{2,2} + D_{2,2}\right)\right] =
m\omega \frac{1}{1 + \frac{2}{N}\left(\frac{\omega}{\Omega_{12}}-1\right)}$
and
$\left(\alpha+\beta+2D'_{2,2}\right) =
m\omega \left[1 + \frac{2}{N}\left(\frac{\Omega_{12}}{\omega}-1\right)\right]$
for the ``$-$'' branch.

Now, expressions (\ref{S_12_4}) can readily be evaluated and the following picture
of higher-order inter-species fragmentation is found:
\beqs\label{S4_INTER_EXAM}
\beqn\label{S4_12_+_EXAM}
& & s_{12,+}^{(4)} = m\omega,
\qquad
\rho_{12,+}^{(4)} = 0,
\qquad
1 - \rho_{12,+}^{(4)} = 1, \
\eeqn
indicating that there is no contribution to fragmentation from the center-of-mass `mixed coordinate' $u_1, u'_1$.
This is additional to the no contribution to fragmentation coming from the relative coordinates
$u_2, u'_2$ and $v_2, v'_2$,
see (\ref{4_RDM_12_SYM_DIAG}).
For the relative center-of-mass `mixed coordinate' $v_1, v'_1$,
on the other end,
one finds
\beqn\label{S4_12_-_EXAM}
& & s_{12,-}^{(4)} = m\omega \sqrt{\frac{1 + \frac{2}{N}\left(\frac{\Omega_{12}}{\omega}-1\right)}
{1 + \frac{2}{N}\left(\frac{\omega}{\Omega_{12}}-1\right)}},
\nonumber \\ \nonumber \\
& & \rho_{12,-}^{(4)} = \frac{\sqrt{\left[1 + \frac{2}{N}\left(\frac{\Omega_{12}}{\omega}-1\right)\right]
\left[1 + \frac{2}{N}\left(\frac{\omega}{\Omega_{12}}-1\right)\right]}-1}
{\sqrt{\left[1 + \frac{2}{N}\left(\frac{\Omega_{12}}{\omega}-1\right)\right]
\left[1 + \frac{2}{N}\left(\frac{\omega}{\Omega_{12}}-1\right)\right]}+1},
\nonumber \\
& & 1 - \rho_{12,-}^{(4)} = \frac{2}
{\sqrt{\left[1 + \frac{2}{N}\left(\frac{\Omega_{12}}{\omega}-1\right)\right]
\left[1 + \frac{2}{N}\left(\frac{\omega}{\Omega_{12}}-1\right)\right]}+1}, \
\eeqn
\eeqs
namely, that the fragmentation of pairs of distinguishable pairs 
fully originates from the relative center-of-mass `mixed coordinate' $v_1, v'_1$.
We see that also the higher-order inter-species fragmentation is governed by the ratio $\frac{\Omega_{12}}{\omega}$
and takes place both at the attractive and repulsive sectors of interactions.
Concluding, higher-order inter-species fragmentation is proved.

Now, one can compute the ratio $\frac{\Omega_{12}}{\omega}=\sqrt{1+\frac{4N}{m\omega^2}\lambda_{12}}$
for which the inter-species reduced four-particle density matrix is $50\%$ fragmented as in (\ref{rho_half}).
Since $\rho_{12,+}^{(4)}=0$ does not contribute in this specific case,
the only contribution to fragmentation comes from
$\rho_{12,-}^{(4)}$.
Thus, solving (\ref{S4_12_-_EXAM}) for $50\%$ distinguishable-four-particle-function fragmentation we obtain
\beqn\label{FRAG_INTER_4}
\rho_{12,-}^{(4)}=\frac{1}{2} \quad
\Longrightarrow \quad
\frac{\Omega_{12}}{\omega} =
\left(1 + \frac{2N^2}{N-2}\right) \pm \sqrt{\left(1 + \frac{2N^2}{N-2}\right)^2 - 1}.
\eeqn
As for distinguishable pairs,
there are two `reciprocate' solutions,
one for strong attractions and the second close to the border of stability for repulsions.
Also, to achieve the same degree of $50\%$
with a larger number $\frac{N}{2}$ of distinguishable four-boson aggregates,
a stronger attraction or repulsion is needed.
Furthermore,
comparing distinguishable-four-boson and distinguishable-two-boson fragmentation at the same $50\%$ value,
one sees from (\ref{FRAG_INTER_4}) and (\ref{FRAG_INTER_2}) that 
slightly weaker interactions, attractions or repulsions, are needed for the former.
This behavior of fragmentation of increasing orders of inter-species reduced density matrices
is analogous to and generalizes
that of intra-species and single-species reduced density matrices,
see the previous section and the appendix, respectively.

Finally, we present for completeness
the inter-species four-particle density,
i.e., the diagonal part
$\rho_{12}^{(4)}(x_1,x_2,y_1,y_2)=
\rho_{12}^{(4)}(x_1,x_2,x'_1=x_1,x'_2=x_2,y_1,y_2,y'_1=y_1,y'_2=y_2)$,
which is given by
\beqn\label{DENSITY_4_12_EXAM}
& &
\rho_{12}^{(4)}(x_1,x_2,y_1,y_2) = N^2 \left(N-1\right)^2
\left(\frac{\alpha-\beta}{\pi}\right)
e^{-\frac{\alpha-\beta}{2}\left(x_1-x_2\right)^2}
e^{-\frac{\alpha-\beta}{2}\left(y_1-y_2\right)^2} \times \nonumber \\
& & \times 
\left[\frac{\alpha+\beta+2\left(C_{2,2}-D_{2,2}\right)}{\pi}\right]^\frac{1}{2}
e^{-\frac{\alpha+\beta+2\left(C_{2,2}-D_{2,2}\right)}{4}\left[\left(x_1+x_2\right)+\left(y_1+y_2\right)\right]^2}
\times \nonumber \\
& & \times 
\left[\frac{\alpha+\beta+2\left(C_{2,2}+D_{2,2}\right)}{\pi}\right]^\frac{1}{2}
e^{-\frac{\alpha+\beta+2\left(C_{2,2}+D_{2,2}\right)}{4}\left[\left(x_1+x_2\right)-\left(y_1+y_2\right)\right]^2} = \nonumber \\
& & 
= N^2 \left(N-1\right)^2 \left(\frac{m\omega}{\pi}\right)^{\frac{3}{2}}
e^{-\frac{m\omega}{2}\left(x_1-x_2\right)^2} e^{-\frac{m\omega}{2}\left(y_1-y_2\right)^2} 
e^{-\frac{m\omega}{4}\left[\left(x_1+x_2\right)+\left(y_1+y_2\right)\right]^2} \times \nonumber \\
& & \times \left(\frac{m\omega}{\pi\left[1 + \frac{2}{N}\left(\frac{\omega}{\Omega_{12}}-1\right)\right]}\right)^{\frac{1}{2}}
e^{-\frac{m\omega}{4\left[1 + \frac{2}{N}\left(\frac{\omega}{\Omega_{12}}-1\right)\right]}
\left[\left(x_1+x_2\right)-\left(y_1+y_2\right)\right]^2}.
\eeqn
To proceed,
the size of the distinguishable four-boson cloud can be estimated 
from the widths of the respective Gaussian functions in the density (\ref{DENSITY_4_12_EXAM}).
Thus, we obtain
\beqs\label{WIDTH_4_INTER}
\beqn\label{WIDTH_4_INTER_GEN}
& & \sigma_{12,\frac{x_1-x_2}{\sqrt{2}}}^{(4)} = \sqrt{\frac{1}{2m\omega}}, \quad
\sigma_{12,\frac{y_1-y_2}{\sqrt{2}}}^{(4)} = \sqrt{\frac{1}{2m\omega}}, \quad
\sigma_{12,\frac{\left(x_1+x_2\right)+\left(y_1+y_2\right)}{2}}^{(4)} =
\sqrt{\frac{1}{2m\omega}}, \nonumber \\
& & 
\sigma_{12,\frac{\left(x_1+x_2\right)-\left(y_1+y_2\right)}{2}}^{(4)} = \sqrt{\frac{1 + \frac{2}{N}\left(\frac{\omega}{\Omega_{12}}-1\right)}{2m\omega}}. \
\eeqn
To show the combined effect of the inter-species and intra-species interactions
accompanying
fragmentation of $\rho_{12}^{(4)}(x_1,x_2,x'_1,x'_2,y_1,y_2,y'_1,y'_2)$,
it is instrumental to compute the sizes (\ref{WIDTH_4_INTER_GEN}) for large inter-species attractions or 
inter-species repulsions at the border of stability.
We obtain, respectively,
\beqn\label{WIDTH_4_INTER_LIM}
& & \!\!\!\!\!\!\!\! \lim_{\frac{\Omega_{12}}{\omega} \to \infty} \sigma_{12,\frac{\left(x_1+x_2\right)-\left(y_1+y_2\right)}{2}}^{(4)} = 
\sqrt{\frac{1 - \frac{2}{N}}{2m\omega}}, \qquad
\sigma_{12,\frac{\left(x_1+x_2\right)-\left(y_1+y_2\right)}{2}}^{(4)}
\longrightarrow \infty \quad \mathrm{for} \quad \frac{\Omega_{12}}{\omega} \to 0^+,
\
\eeqn
\eeqs
where
$\sigma_{12,\frac{x_1-x_2}{\sqrt{2}}}^{(4)}$,
$\sigma_{12,\frac{y_1-y_2}{\sqrt{2}}}^{(4)}$, and
$\sigma_{12,\frac{\left(x_1+x_2\right)+\left(y_1+y_2\right)}{2}}^{(4)}$
do not depend on the interactions.
We see that the size of the inter-species four-boson density
saturates as well at about the trap's size and does not depend on the strengths of interactions
in the limit of strong inter-species attractions.
As for the pair of distinguishable bosons,
a strong fragmentation is possible in the mixture with
hardy any shrinking of the density in comparison with the bare trap
due to the condition $\lambda+\lambda_{12}=0$,
i.e., that strong inter-species attractive interaction is accompanied by 
strong intra-species repulsion of equal magnitude.
Summarizing, inter-species fragmentation in higher-order reduced density matrices
is also governed by the ratio $\frac{\Omega_{12}}{\omega}$
and takes place both at the attractive and repulsive sectors of interactions.

\subsection{Inter-species entanglement and the limit of an infinite number of particles}\label{MORE_Schmidt}

In the previous sections the reduced density matrices 
for identical and distinguishable pairs of bosons were diagonalized and
the intra-species and inter-species fragmentations explored.
Both kinds of fragmentations are critical phenomena in the sense that,
going to the limit of an infinite number of particles while keeping the interaction parameters 
(products of the number of particles times the interaction strengths) constant,
the respective reduced density matrix per particle becomes $100\%$ condensed \cite{HIM_MIX_RDM}.
This can be easily obtained from the leading natural eigenvalues of the natural functions
explicitly obtained above,
see the general (\ref{S_1_1}), (\ref{S_1_2}), (\ref{S_12_2}), (\ref{S_12_4})
and specific (\ref{S_INTRA_EXAM}), (\ref{S2_INTER_EXAM}),  (\ref{S4_INTER_EXAM}) expressions,
which are all equal to $1$ in this limit.

In the present, concluding subsection we touch upon a property of the mixture which does not diminish at the limit
of an infinite number of particles.
Classifying properties of Bose-Einstein condensates and their mixtures
at the limit of an infinite number of particles,
and especially when many-body and mean-field theories do not coincide,
is an active field of research,
where variances and the overlap between the many-body and mean-field
wavefunctions are discussed elsewhere,
see \cite{HIM_MIX_VAR,HIM_MIX_FLOQUET,HIM_MIX_CP,INF1,INF2,INF3,INF4,INF5,INF6,INF7,INF8,INF9}.
Here, combining the techniques used in the previous sections,
we apply Mehler's formula to perform the Schmidt decomposition of the wavefunction.

Let us examine the mixture's wavefunction,
for which the coordinates of the two species are coupled to each other owing to the inter-species interaction,
see the last term in (\ref{HIM_MIX_WF_DEN_MAT1}).
To remind, the wavefunction is obtained by representing the Hamiltonian (\ref{HIM_MIX}) with the mixture's Jacoby coordinates,
for which it is fully diagonalized, and translating back to the laboratory frame.
To decouple the coordinates of each species, in the sense of prescribing the Schmidt decomposition of the wavefunction,
it is useful to go `half a step' backward,
and express (\ref{HIM_MIX_WF_DEN_MAT1}) using the individual species' Jacoby coordinates.

The Jacoby coordinates of each species are given by
\beqn\label{JACOBY_12}
& & X_k = \frac{1}{\sqrt{k(k+1)}} \sum_{j=1}^k \left(x_{k+1}-x_j\right), \quad 1\le k \le N-1, \qquad
X_{N} = \frac{1}{\sqrt{N}} \sum_{j=1}^{N} x_j, \nonumber \\
& & Y_k = \frac{1}{\sqrt{k(k+1)}} \sum_{j=1}^k \left(y_{k+1}-y_j\right), \quad 1\le k \le N-1, \qquad
Y_{N} = \pm \frac{1}{\sqrt{N}} \sum_{j=1}^{N} y_j, \
\eeqn
where, for the derivation given below,
it is useful to distinguish between the two cases for the definition of, say, $Y_N$:
The plus sign is assigned to positive $\gamma$, namely,
to attractive inter-species interactions for which $\Omega_{12} > \omega$,
and the minus sign is assigned to negative $\gamma$, i.e.,
to repulsive inter-species interactions where $\Omega_{12} < \omega$.

For the symmetric mixture,
given the above Jacobi coordinates of each species, Eq.~(\ref{JACOBY_12}),
the wavefunction reads
\beqs\label{WF_MIX}
\beqn\label{WF_MIX_JACOBY}
& & \Psi(X_1,\ldots,X_{N},Y_1,\ldots,Y_{N}) =
\left(\frac{m\Omega}{\pi}\right)^{\frac{N-1}{2}}
\left(\frac{M_{12}\Omega_{12}}{\pi}\right)^{\frac{1}{4}}
\left(\frac{M\omega}{\pi}\right)^{\frac{1}{4}} \times \nonumber \\
& & \times e^{-\frac{1}{2} m \Omega \sum_{k=1}^{N-1} \left(X_k^2 + Y_k^2\right)}
e^{-\frac{\frac{1}{2}m\left(\Omega_{12}+\omega\right)}{2}\left(X_{N}^2 + Y_{N}^2\right)}
e^{\pm\frac{1}{2}m\left(\Omega_{12}-\omega\right)X_{N}Y_{N}}. \
\eeqn
Indeed, all relative coordinates are decoupled and the only coupling due to the inter-species interaction
is between the center-of-mass
$X_N$ of species $1$ bosons
and the center-of-mass
$Y_N$ of species $2$ bosons.
Consequently, applying Mehler's formula to the intra-species
center-of-mass Jacoby coordinates $X_{N}$ and $Y_{N}$
the Schmidt decomposition of (\ref{WF_MIX_JACOBY}) is readily performed and given by
\beqn\label{WF_MIX_SCHMIDT}
& & \Psi(X_1,\ldots,X_{N},Y_1,\ldots,Y_{N})
= \sum_{n=0}^{\infty} \sqrt{1-\rho^2_{SD}} \rho_{SD}^n \Phi_{1,n}(X_1,\ldots,X_N) \Phi_{2,n}(Y_1,\ldots,Y_N), \nonumber \\
& & \Phi_{1,n}(X_1,\ldots,X_N) = \left(\frac{m\Omega}{\pi}\right)^{\frac{N-1}{4}}
e^{-\frac{1}{2} m \Omega \sum_{k=1}^{N-1} X_k^2}
\frac{1}{\sqrt{2^n n!}} \left(\frac{s_{SD}}{\pi}\right)^{\frac{1}{4}} H_n\left(\sqrt{s_{SD}}X_N\right) e^{-\frac{1}{2}s_{SD} X_N^2}, \nonumber \\
& & \Phi_{2,n}(Y_1,\ldots,Y_N) = \left(\frac{m\Omega}{\pi}\right)^{\frac{N-1}{4}}
e^{-\frac{1}{2} m \Omega \sum_{k=1}^{N-1} Y_k^2}
\frac{1}{\sqrt{2^n n!}} \left(\frac{s_{SD}}{\pi}\right)^{\frac{1}{4}} H_n\left(\sqrt{s_{SD}}Y_N\right) e^{-\frac{1}{2}s_{SD} Y_N^2}, \nonumber \\
& & \sqrt{1-\rho^2_{SD}} =
\frac{2\sqrt{\frac{\Omega_{12}}{\omega}}}{1+\frac{\Omega_{12}}{\omega}}, \qquad
\rho_{SD} =
\frac{\left(\frac{\Omega_{12}}{\omega}\right)^{\pm 1}-1}{\left(\frac{\Omega_{12}}{\omega}\right)^{\pm 1}+1}, \qquad
s_{SD} = m\sqrt{\omega\Omega_{12}}.
\
\eeqn
\eeqs
We remind that the plus sign is for attraction and the minus for repulsion,
which is what guarantees that $\rho_{SD}$ 
and consequently the Schmidt coefficients $\sqrt{1-\rho^2_{SD}} \rho_{SD}^n$, $n=0,1,2,3,\ldots$
are always positive.
$s_{SD}$ defines the inverse width of the individual species' center-of-mass Gaussians in
the Schmidt basis $\Phi_{1,n}(X_1,\ldots,X_N)$ and $\Phi_{2,n}(Y_1,\ldots,Y_N)$. 

Let us concisely discuss properties of the Schmidt decomposition of the mixture, Eq.~(\ref{WF_MIX_SCHMIDT}). 
Clearly and interestingly,
the Schmidt coefficients are independent of the intra-species dressed frequency $\Omega$,
which only appears in conjunction with intra-species relative coordinates, 
i.e., the Schmidt coefficients depend solely on the
inter-species interaction.
Furthermore,
there is a kind of symmetry between respective attractive and repulsive inter-species interactions,
as one gets the
same Schmidt coefficients for the inter-species frequency $\frac{\Omega_{12}}{\omega}=\sqrt{1+\frac{4N\lambda_{12}}{m\omega^2}}$ and
inverse frequency $\frac{\omega}{\Omega_{12}}=\frac{1}{\sqrt{1+\frac{4N\lambda_{12}}{m\omega^2}}}$.

Last but not least,
the same Schmidt coefficients are obtained when the product
of the number of bosons in each species times the inter-species interaction strength, $N\lambda_{12}$, is held fixed,
and $N$ is increased to infinity.
In other words,
whereas identical and distinguishable bosons, pairs, four-particle aggregates, etc.
are $100\%$ condensed at the limit of an infinite number of particles, 
i.e., the leading eigenvalue of 
all finite-order intra-species and inter-species reduced density matrices per particle is $1$,
the mixture's wavefunction exhibits a fixed amount of entanglement at the infinite-particle-number limit.
This is a good place to bring the present study to an end. 

\subsection{Summary and Outlook}\label{SUM_OUT}

The present work aims at developing and combining concepts from quantum theory
of many-particle systems with novel results on the physics
of trapped mixtures of Bose-Einstein condensates.
The notions of natural orbitals and
natural geminals are fundamental to many-particle systems
made of identical particles.
These natural functions entail the diagonalization of the reduced one-particle and two-particle density matrices, respectively.
In a mixture of two kinds of identical particles,
here explicitly two types of bosons,
there are, naturally, identical bosons and pairs made of indistinguishable bosons of either species.
To find their natural orbitals and natural geminals,
the construction and subsequent diagonalization of
respective intra-species reduced density matrices is in need.
In the mixture there are, additionally, pairs made of distinguishable bosons.
Analogously, their theoretical description would require
assembling, diagonalizing, and analyzing the inter-species
reduced two-particle density matrix. 
In the present work we have investigated pairs made of identical or distinguishable
bosons in a mixture of Bose-Einstein condensates,
covering both the structure of the respective natural pair functions,
on the more formal theoretical side,
and the exploration of pairs' fragmentation.
Like identical bosons,
which can, depending on whether the reduced one-particle density matrix has
one or more macroscopic eigenvalues,
be condensed or fragmented,
so do pairs of bosons.
We showed in the present work that,
in the mixture, both pairs made of identical bosons
and pairs consisting of distinguishable bosons
can be condensed and more so fragmented.

To tackle the above and other questions,
we employed a solvable model,
the symmetric harmonic-interaction model for mixtures.
The natural geminals for pairs made of identical or distinguishable bosons
were explicitly contracted as a function of the inter-species and intra-species interactions.
This was done
by diagonalizing the corresponding intra-species and inter-species reduced two-particle density matrices
using applications of Mehler's formula on appropriately-constructed linear combinations of intra-species and inter-species coordinates.
Here, the role of the mixture's center-of-mass and relative center-of-mass coordinates was identified and explained.
The structure of identical and distinguishable pairs in the mixture was discussed,
and a generalization to pairs of distinguishable pairs using the inter-species reduced four-body density matrix was made.
A particular case, where attractive and repulsive inter-species and intra-species interactions are opposite in magnitude,
has been worked out explicitly.
Fragmentation of bosons, pairs, and pairs of pairs in the mixture has been proven,
and the size of the respective densities analyzed.
Last but not least,
as a complementary investigation,
the exact Schmidt decomposition of the mixture's wavefunction was performed.
The entanglement between the two species was shown to be governed by the coupling of their individual center-of-mass coordinates
and, consequently, not to vanish at the limit of an infinite number of particles where
any finite-order intra-species and inter-species reduced density matrix per particle is 100\% condensed.

The present investigations suggest several directions for further developments.
We have treated the symmetric mixture
and an anticipated extension to
generic trapped mixtures, with different numbers of bosons, masses, and interaction strengths for each species, would be in place.
In what capacity the fragmentation of identical pairs in the different species can be made to differ,
and to what extent the fragmentation of distinguishable pairs would become more complex? 
Do the center-of-mass and relative center-of-mass coordinates keep their
role in the diagonalization of inter-species reduced density matrices for a generic mixture?
It also makes sense, in a generic mixture,
to investigate fragmentation of
aggregates with unequal numbers of bosons from each species,
like, for instance, the analysis of inter-species reduced three-particle density matrices.
Another extension foreseen is to mixtures with more species and, if feasible,
to generic multi-species mixtures where, e.g., one species could serve as a bridge between two baths.
Finally, one could forecast that the topic of Bose-Einstein condensates and mixtures in the limit of an infinite number of particles
would be enriched by exploring the Schmidt decomposition of the wavefunction.
Recall that at the infinite-particle-number limit any finite-order intra-species and inter-species reduced
density matrix per particle is $100\%$ condensed.
Here, observables' variances and wavefunctions' overlaps
have deepened our understanding of the
differences between many-body and mean-field theories of Bose-Einstein condensates and mixtures
at limit of an infinite number of particles,
but are properties already defined for single-species bosons.
The Schmidt decomposition, on the other hand, is a property that
enters the topic of the infinite-particle-number limit starting, obviously,
only from a two-species mixture.
All of which paves the way for further intriguing investigations to come.
    
\section*{Acknowledgements}

This research was supported by the Israel Science Foundation 
(Grant No. 1516/19). 

\appendix

\section{Comparison to fragmentation in the single-species system}\label{APP}

The Hamiltonian of the single-species harmonic-interaction model is given by \cite{HIM_RDM}
\beq\label{HIM_SIN}
\hat H(x_1,\ldots,x_N) = \sum_{j=1}^N \left( -\frac{1}{2m}\frac{\partial^2}{\partial x_j^2} + \frac{1}{2}m\omega^2 x_j^2 \right)
+ \lambda \sum_{1\le j <k}^N \left(x_j-x_k\right)^2. 
\eeq
Employing single-species Jacoby coordinates
and translating back to the laboratory frame,
the $N$-boson wavefunction and corresponding density matrix are given by
\beqn\label{HIM_SIN_WF_DEN}
& & 
\Psi(x_1,\ldots,x_N) = \left(\frac{m\Omega}{\pi}\right)^{\frac{N-1}{4}} \left(\frac{m\omega}{\pi}\right)^{\frac{1}{4}}
e^{-\frac{\alpha}{2}\sum_{j=1}^N x_j^2 - \beta \sum_{1\le j < k}^N x_jx_k}, \nonumber \\
& &
\Psi(x_1,\ldots,x_N)\Psi^\ast(x'_1,\ldots,x'_N) = 
 \left(\frac{m\Omega}{\pi}\right)^{\frac{N-1}{2}} \left(\frac{m\omega}{\pi}\right)^{\frac{1}{2}}
e^{-\frac{\alpha}{2}\sum_{j=1}^N \left(x_j^2 + {x'_j}^2\right)
- \beta \sum_{1\le j < k}^N \left(x_jx_k + x'_jx'_k\right)}, \nonumber \\
& &
\alpha
= m\Omega + \beta = m\Omega\left[1+\frac{1}{N}\left(\frac{\omega}{\Omega}-1\right)\right], \quad
\beta = m\Omega\frac{1}{N}\left(\frac{\omega}{\Omega}-1\right), \quad
\Omega = \sqrt{\omega^2+\frac{2\lambda N}{m}}. \
\eeqn
The stability of the system means that the
interaction satisfies $\lambda > - \frac{m\omega^2}{2N}$.

The reduced one-particle density matrix reads
\beqn\label{1_SIN_RDM}
& &
\rho^{(1)}(x,x')
= N \left(\frac{\alpha+C_1}{\pi}\right)^{\frac{1}{2}}
e^{-\frac{\alpha+\frac{C_1}{2}}{2}\left(x^2+{x'}^2\right)} 
e^{- \frac{1}{2} C_1 xx'}, \nonumber \\
& &
\alpha + C_1 = (\alpha-\beta) \frac{(\alpha-\beta)+N\beta}{(\alpha-\beta)+(N-1)\beta} =
m\Omega\frac{1}{1+\frac{1}{N}\left(\frac{\Omega}{\omega}-1\right)}.
\eeqn
Comparing the structure of reduced single-particle density matrix with
that of Mehler's formula \cite{HIM_JCP,HIM_SCH} one readily has
\beqn\label{SIN_FRAG_1}
& & s^{(1)} = \sqrt{\alpha\left(\alpha+C_1\right)} =
m\Omega\sqrt{\frac{1+\frac{1}{N}\left(\frac{\omega}{\Omega}-1\right)}{1+\frac{1}{N}\left(\frac{\Omega}{\omega}-1\right)}},
\nonumber \\
& & \rho^{(1)} = \frac{\alpha - s^{(1)}}{\alpha + s^{(1)}} =
\frac{\sqrt{\left[1+\frac{1}{N}\left(\frac{\omega}{\Omega}-1\right)\right]\left[1+\frac{1}{N}\left(\frac{\Omega}{\omega}-1\right)\right]}-1}{\sqrt{\left[1+\frac{1}{N}\left(\frac{\omega}{\Omega}-1\right)\right]\left[1+\frac{1}{N}\left(\frac{\Omega}{\omega}-1\right)\right]}+1},
\nonumber \\
& & 1 - \rho^{(1)} = \frac{2s^{(1)}}{\alpha +s^{(1)}} = \frac{2}{\sqrt{\left[1+\frac{1}{N}\left(\frac{\omega}{\Omega}-1\right)\right]\left[1+\frac{1}{N}\left(\frac{\Omega}{\omega}-1\right)\right]}}. \
\eeqn

The reduced two-particle density matrix $\rho^{(2)}(x_1,x_2,x'_1,x'_2)$ reads, after rotation of coordinates,
\beqn\label{2_RDM_SIN_DIAG}
& &
\rho^{(2)}(q_1,q'_1,q_2,q'_2) = 
N(N-1)
\left(\frac{\alpha-\beta}{\pi}\right)^{\frac{1}{2}}
e^{-\frac{\alpha-\beta}{2}\left(q_2^2 + {q'_2}^2\right)} \times \nonumber \\
& & \times
\left(\frac{\alpha+\beta+2C_2}{\pi}\right)^{\frac{1}{2}}
e^{-\frac{\alpha+\beta+C_2}{2}\left(q_1^2 + {q'_1}^2\right)}
e^{-C_2 q_1q'_1}, \nonumber \\
& &
\alpha + \beta + 2C_2 = (\alpha-\beta) \frac{(\alpha-\beta)+N\beta}{(\alpha-\beta)+(N-2)\beta}
=m\Omega\frac{1}{1+\frac{2}{N}\left(\frac{\Omega}{\omega}-1\right)},
\eeqn
where $q_1 = \frac{1}{\sqrt{2}}\left(x_1+x_2\right)$, $q_2 = \frac{1}{\sqrt{2}}\left(x_1-x_2\right)$
and
$q'_1 = \frac{1}{\sqrt{2}}\left(x'_1+x'_2\right)$, $q'_2 = \frac{1}{\sqrt{2}}\left(x'_1-x'_2\right)$.
Comparing the structure of the reduced two-particle density matrix with that of Mehler's formula (\ref{MEHLER})
we readily find
\beqn\label{SIN_FRAG_2}
& & s^{(2)} = \sqrt{(\alpha+\beta)\left(\alpha+\beta+2C_2\right)} =
m\Omega\sqrt{\frac{1+\frac{2}{N}\left(\frac{\omega}{\Omega}-1\right)}{1+\frac{2}{N}\left(\frac{\Omega}{\omega}-1\right)}},
\nonumber \\
& & \rho^{(2)} = \frac{(\alpha+\beta) - s^{(2)}}{(\alpha+\beta) + s^{(2)}} =
\frac{\sqrt{\left[1+\frac{2}{N}\left(\frac{\omega}{\Omega}-1\right)\right]\left[1+\frac{2}{N}\left(\frac{\Omega}{\omega}-1\right)\right]}-1}{\sqrt{\left[1+\frac{2}{N}\left(\frac{\omega}{\Omega}-1\right)\right]\left[1+\frac{2}{N}\left(\frac{\Omega}{\omega}-1\right)\right]}+1},
\nonumber \\
& & 1 - \rho^{(2)} = \frac{2s^{(2)}}{(\alpha+\beta) +s^{(2)}} = \frac{2}{\sqrt{\left[1+\frac{2}{N}\left(\frac{\omega}{\Omega}-1\right)\right]\left[1+\frac{2}{N}\left(\frac{\Omega}{\omega}-1\right)\right]}}, \
\eeqn
where $\alpha+\beta = m\Omega\left[1+\frac{2}{N}\left(\frac{\omega}{\Omega}-1\right)\right]$.
We see from (\ref{SIN_FRAG_1}) and (\ref{SIN_FRAG_2})
that fragmentation of bosons and pairs is governed, 
in the single-species harmonic-interaction model,
by the ratio $\frac{\Omega}{\omega}$
and takes place both at the attractive and repulsive sectors of the interaction.

Similarly to the main text,
we compute for which ratio $\frac{\Omega}{\omega}$,
or, equivalently, for which interaction
$\lambda = \frac{m\omega^2}{2N}\left[\left(\frac{\Omega}{\omega}\right)^2-1\right]$,
the two-particle and one-particle reduced density matrices are macroscopically fragmented as in (\ref{rho_half}).
Note the difference that, here, $\frac{\Omega}{\omega}$ is the single-species interaction
and in the main text, for the mixtue, $\frac{\Omega_{12}}{\omega}$ is the inter-species interaction.
Thus, solving (\ref{SIN_FRAG_1}) for $50\%$ natural-orbital fragmentation we find
\beqn\label{FRAG_INTRA_SIG_1}
 \rho^{(1)}=\frac{1}{2} \quad
\Longrightarrow \quad
\frac{\Omega}{\omega} =
\left(1 + \frac{4N^2}{N-1}\right) \pm \sqrt{\left(1 + \frac{4N^2}{N-1}\right)^2 - 1},
\eeqn
and working out (\ref{SIN_FRAG_2}) for $50\%$ natural-geminal fragmentation we get
\beqn\label{FRAG_INTRA_SIG_2}
 \rho_1^{(2)}=\frac{1}{2} \quad
\Longrightarrow \quad
\frac{\Omega}{\omega} = 
\sqrt{1+\frac{2N\lambda}{m\omega^2}} = 
\left(1 + \frac{2N^2}{N-2}\right) \pm \sqrt{\left(1 + \frac{2N^2}{N-2}\right)^2 - 1}.
\eeqn
There are two `reciprocate' solutions for both
the natural geminals and natural orbitals:
Indeed, $50\%$ fragmentation occurs for strong attractions, namely, when $\frac{\Omega}{\omega}$ is large,
or in the vicinity of the border of stability for repulsions, i.e., when $\frac{\Omega}{\omega}$ is close to zero.
Also, to achieve the same degree of $50\%$ with a larger number $N$ of bosons,
a stronger attraction or repulsion is needed.
Finally, comparing natural-geminal with natural-orbital fragmentation at the same $50\%$ value,
one sees from (\ref{FRAG_INTRA_SIG_2}) and (\ref{FRAG_INTRA_SIG_1}) that slightly weaker
interactions, attractions or repulsions, are needed for the former,
in a similar manner to intra-species fragmentation in the mixture
discussed in the main text.

Finally, we prescribe the one-particle and two-particle densities, i.e.,
the diagonal parts
$\rho^{(1)}(x)=\rho^{(1)}(x,x'=x)$ and
$\rho^{(2)}(x_1,x_2)=\rho^{(2)}(x_1,x_2,x'_1=x_1,x'_2=x_2)$
which read
\beqn\label{DENSITIES_1_2_SIN_EXAM}
& & \rho^{(1)}(x) =
N \left(\frac{\alpha+C_{1}}{\pi}\right)^{\frac{1}{2}} e^{-\left(\alpha+C_{1}\right)x^2} = 
N \left(\frac{m\Omega}{\pi\left[1 + \frac{1}{N}\left(\frac{\Omega}{\omega}-1\right)\right]}\right)^{\frac{1}{2}}
e^{-\frac{m\Omega}{1 + \frac{1}{N}\left(\frac{\Omega}{\omega}-1\right)}x^2},
\nonumber \\
& & \rho^{(2)}(x_1,x_2) = 
N(N-1)
\left(\frac{\alpha-\beta}{\pi}\right)^{\frac{1}{2}}
e^{-\frac{\alpha-\beta}{2}\left(x_1-x_2\right)^2}
\left(\frac{\alpha+\beta+2C_{2}}{\pi}\right)^{\frac{1}{2}}
e^{-\frac{\alpha+\beta+2C_{2}}{2}\left(x_1+x_2\right)^2} = \nonumber \\
& &
= N(N-1)
\left(\frac{m\Omega}{\pi}\right)^{\frac{1}{2}}
e^{-\frac{m\Omega}{2}\left(x_1-x_2\right)^2}
\left(\frac{m\Omega}{\pi\left[1 + \frac{2}{N}\left(\frac{\Omega}{\omega}-1\right)\right]}\right)^{\frac{1}{2}}
e^{-\frac{m\Omega}{2\left[1 + \frac{2}{N}\left(\frac{\Omega}{\omega}-1\right)\right]}\left(x_1+x_2\right)^2}.
\eeqn
Here, in the single-species case, $\lambda$ governs the size of the densities which, as we now show,
depend on the interaction stronger than for the mixture,
also see the main text for comparison and further discussion.

We can deduce
the size
of pairs' and bosons' clouds using the widths of the
corresponding Gaussian functions
in the densities (\ref{DENSITIES_1_2_SIN_EXAM}). 
Hence, we have
\beqs\label{WIDTH_INTRA_SIN}
\beqn\label{WIDTH_INTRA_SIN_GEN}
& &
\sigma_{x}^{(1)} = \sqrt{\frac{1 + \frac{1}{N}\left(\frac{\Omega}{\omega}-1\right)}{2m\Omega}}, \nonumber \\
& &
\sigma_{\frac{x_1+x_2}{\sqrt{2}}}^{(2)} = \sqrt{\frac{1 + \frac{2}{N}\left(\frac{\Omega}{\omega}-1\right)}{2m\Omega}},
\quad
\sigma_{\frac{x_1-x_2}{\sqrt{2}}}^{(2)} = \sqrt{\frac{1}{2m\Omega}}. \
\eeqn
To describe further effects of the interaction $\lambda$
accompanying fragmentation of the reduced density matrices (\ref{SIN_FRAG_1}) and (\ref{SIN_FRAG_2}),
we compute the sizes (\ref{WIDTH_INTRA_SIN_GEN}) for strong attractions or
repulsions at the border of stability.
We obtain, respectively,
\beqn\label{WIDTH_INTRA_SIN_LIM}
& & \lim_{\frac{\Omega}{\omega} \to \infty} \sigma_{x}^{(1)} = 
\sqrt{\frac{1}{2m\omega N}}, \qquad
\sigma_{x}^{(1)} \longrightarrow \infty \quad \mathrm{for} \quad \frac{\Omega}{\omega} \to 0^+, \nonumber \\
& & \lim_{\frac{\Omega}{\omega} \to \infty} \sigma_{\frac{x_1+x_2}{\sqrt{2}}}^{(2)} = 
\sqrt{\frac{1}{m\omega N}}, \qquad
\sigma_{\frac{x_1+x_2}{\sqrt{2}}}^{(2)} \longrightarrow \infty \quad \mathrm{for} \quad \frac{\Omega}{\omega} \to 0^+, \nonumber \\
& & \lim_{\frac{\Omega}{\omega} \to \infty} \sigma_{\frac{x_1-x_2}{\sqrt{2}}}^{(2)} = 0, \qquad
\sigma_{\frac{x_1-x_2}{\sqrt{2}}}^{(2)} \longrightarrow \infty \quad \mathrm{for} \quad \frac{\Omega}{\omega} \to 0^+,
\
\eeqn
\eeqs
as all widths (\ref{WIDTH_INTRA_SIN_GEN}) depend on the interaction strength.
The size of the densities 
for strong attractions 
diminishes to much smaller values than the trap's size that do not depend on the strength of interaction.
Thus, a high degree of fragmentation due to
the strong attractive interaction
is possible in the single-species system only together with shrinking of the density.
Alternatively, towards the edge of stability the density of the single-species system
expands due to repulsive interaction unlimitedly as the degree of fragmentation increases.
These should be compared to
and contrasted with the results in the main text 
for the intra-species densities and 
the interplay between inter-species and intra-species interactions
within intra-species fragmentation
in the mixture.

\end{document}